%Paper: hep-th/9204074
%From: J.M.C.F.Govaerts%durham.ac.uk@ib.rl.ac.uk
%Date: Thu, 23 Apr 92 12:20:05 BST
%Date (revised): Thu, 23 Jul 92 16:57:37 BST

%%%%%%%%%%%%%%%%%%%%%%%%%%%%%%%
%% This is a Plain TeX file  %%
%%%%%%%%%%%%%%%%%%%%%%%%%%%%%%%
\def\pd{\partial}
\def\ts{\thinspace}
\magnification=1200
\pageno=0
\parskip 3 pt plus 1pt minus 1 pt
\rightline{DTP-92/17}
\rightline{April 1992}
\vskip 2 true cm
\centerline{EULER HIERARCHIES AND UNIVERSAL EQUATIONS}
\vskip 2.5 true cm
\centerline{D.B. FAIRLIE and J. GOVAERTS}
\vskip 0.5 true cm
\centerline{\it{Department of Mathematical Sciences}}
\centerline{\it{University of Durham, Durham DH1 3LE, England}}
\vskip 2 true cm
\centerline{Abstract}
\vskip 1 true cm

Finite Euler hierarchies of field theory Lagrangians leading to universal
equations of motion for new types of string and membrane theories and for
{\it classical} topological field theories are constructed. The analysis uses
two main ingredients. On the one hand, there exists a generic finite Euler
hierarchy for one field leading to a universal equation which generalises the
Plebanski equation of self-dual four dimensional gravity. On the other hand,
specific maps are introduced between field theories which provide a
``triangular duality'' between certain classes of arbitrary field theories,
classical topological field theories and generalised string and membrane
theories. The universal equations, which derive from an infinity of
inequivalent Lagrangians, are generalisations of certain reductions of the
Plebanski and KdV equations, and could possibly define new integrable systems,
thus in particular integrable membrane theories. Some classes of solutions
are constructed in the general case. The general solution to some of the
universal equations is given in the simplest cases.

\vfill\eject
\vskip 10pt
\leftline{\bf 1. Introduction}
\vskip 20pt
In two recent papers$\sp {[1,2]}$, the first with A. Morozov,
hierarchies of Lagrangian field theories with the following
properties were introduced.
\item{i)} In any of these hierarchies, the Lagrangian at any given level
-- except of course at the first level -- is essentially proportional to the
equations of motion of the Lagrangian at the previous level (hence the name
Euler hierarchies).
\item{ii)} The proportionality factor mentioned in i) is essentially the very
first Lagrangian in the hierarchy.
\item{iii)} In any of these hierarchies, Lagrangians depend on fields only
through their first and second derivatives, but {\it not\/} on
derivatives of {\it higher\/} order nor on the fields themselves. The first
Lagrangian only depends on first derivatives of the fields. The dependence of
each of the other Lagrangians on second derivatives is
multilinear, and of order equal to the number of times an equation
of motion has been taken to reach that level in the hierarchy.
\item{iv)} All these hierarchies are {\it finite\/}, {\it i.e.} the
iterative procedure implied by i) -- iii) terminates after a {\it finite\/}
number of steps.
\item{v)} For each hierarchy, the last non trivial equations of motion
are universal, namely, up to a
factor, they are {\it independent\/} of the initial Lagrangian out of
which the hierarchy is constructed. The associated infinite number of
conserved charges -- corresponding to the freedom in the choice of
initial Lagrangian -- suggests the possible integrability of these
universal equations (equations of motion are indeed always current
conservation equations for Lagrangians without an explicit dependence
on fields).

\vskip 20pt
Specifically, hierarchies with these properties were shown$\sp {[1,2]}$
to exist in the following cases:
\item{1)} a single field $\phi\ts$ in $d\ts$ dimensions, with an arbitrary
initial Lagrangian (function of first derivatives only)$\sp {[1]}$,
\item{2)} a single field $\phi\ts$ in $(d+1)$ dimensions, the initial
Lagrangian now being an arbitrary homogeneous weight one function of its
arguments$\sp {[1]}$,
\item{3)} $(d+1)$ fields $\phi\sp a\ts$ in $d\ts$ dimensions, with an
arbitrary reparametrisation invariant initial Lagrangian$\sp {[2]}$.

\vskip 10pt
The hierarchies associated with these three cases terminate after $d\ts$
steps, with the following universal equations:
\item{}Case 1)
$$\det{\phi_{ij}} = 0,\eqno(1.1)$$
\item{}Case 2)
$$\det\pmatrix{0&\phi_j\cr\phi_i&\phi_{ij}\cr} = 0,\eqno(1.2)$$
\item{}Case 3)
$$\det({J_a\phi\sp a_{ij}}) = 0.\eqno(1.3)$$

Here, $\phi_i$ and $\phi_{ij}$ denote the partial derivatives
$({\pd\phi}/{\pd x_i})\ts, ({\pd\sp 2\phi}/{(\pd x_i\pd x_j)})$ of the field
$\phi\ts$ with respect to the $d\ts$ or $(d+1)$ coordinates $x_i$. (The same
applies of course to the fields $\phi\sp a$, and obviously the indices $i$ and
$j$ in the equations above refer to lines and columns respectively of the
corresponding matrices. The usual summation convention over repeated indices
is assumed throughout). In (1.3), the quantities $J_a$ are the Jacobians
$$J_a =
(-1)\sp d\epsilon_{ab_1b_2\cdots b_d}\phi\sp {b_1}_1\phi\sp {b_2}_2\dots
\phi\sp {b_d}_d.\eqno(1.4)$$

Eq.(1.1) is a generalisation to $d\ts$ dimensions of a particular reduction
(corresponding to (1.1) for $(d=2)$) of the Plebanski equation$\sp {[3]}$ for
self-dual gravity in four dimensions, which is well known to be an integrable
system. Eq.(1.2) is a generalisation of the original two dimensional Bateman
equation$\sp {[4,5]}$ (corresponding to (1.2) with $(i=1,2)$) which is also
known to be integrable$\sp {[5,1]}$. Finally, (1.3) is a generalisation to a
$(d-1)$-dimensional membrane in a $(d+1)$-dimensional spacetime of the
(universal) equation of motion for a parametrised particle in a flat two
dimensional spacetime (corresponding to $(d=1)$), the latter clearly being
also integrable. Note that (1.3) includes a universal equation for a string
theory in three dimensions ($d=2$), and a universal equation for a membrane
theory in four dimensions ($d=3$).

Remarkably, the three classes of universal equations above are invariant under
arbitrary linear $GL(n)$ transformations in the variables $x_i$ as
well as in the fields $\phi\sp a$, even though neither the initial nor the
successive Lagrangians in the corresponding hierarchies would generally
possess these symmetries. The equations above thus provide examples of
equations of motion admitting an {\it infinite\/} number of Lagrangians,
with symmetry properties that these Lagrangians need not possess!

Moreover, in cases 2) and 3), the properties required of the initial
Lagrangians extend to the whole of the resulting hierarchies. Namely, in case
3), reparametrisation invariance in the coordinates $x_i$ for the field
theory defined by the initial Lagrangian ensures$\sp {[2]}$ the same
invariance property
for {\it all\/} field theories defined by the hierarchy. Thus in particular,
the universal equation (1.3) is reparametrisation covariant in the variables
$x_i$, as is desirable of any string or membrane theory. In case 2), weight
one homogeneity of the initial Lagrangian ensures$\sp {[1]}$ that {\it all\/}
equations of motion of the hierarchy are invariant under arbitrary
redefinitions of the field $\phi$. Thus in particular, the universal Bateman
equation (1.2) possesses this property of general covariance in the field
$\phi\ts$. This notion of classical general
covariance of equations of motion under arbitrary field redefinitions,
{\it i.e.} general covariance on the space of solutions, may obviously be
extended to the case of many fields $\phi\sp a$. Identifying these fields
with coordinates in a target space, field theories leading to such generally
covariant equations of motion provide a class of theories halfway between
ordinary field theories and {\it quantum\/} topological field
theories$\sp {[6]}$. Namely, such theories correspond to {\it classical\/}
topological field theories whose space of classical solutions falls into
diffeomorphic topological classes of the possible target manifolds
parametrised by the fields $\phi\sp a$.

Actually, a generally covariant generalisation of the Bateman equation
(1.2) to many fields was given in Ref.[1], when the number of
fields $\phi\sp a$ is less than the number of coordinates $x_i$. Moreover,
these equations were conjectured$\sp {[1]}$ to be universal and to follow
from a hierarchy construction of the type described in i) -- v) above.
In compact form, these generalised Bateman equations are given by
$$\det\pmatrix{0&\phi\sp a_j\cr\phi\sp b_i&\phi\sp c_{ij}\lambda\sp c\cr} = 0
\  .\eqno(1.5)$$
Here, the determinant is that of a $(D+d\ts)\times (D+d\ts)$ matrix, with
$D\ts$ being the number of fields $\phi\sp a$ and $d\ts$ the number of
coordinates $x_i$ ($D < d\ts$), and the indices $(a,i)$ and $(b,j)$ refer
to lines and columns respectively (in particular, the entry ``$\ts0\ts$''
stands for the $D\times D$ null matrix). The quantities $\lambda\sp c$ are
arbitrary coefficients, in terms of which the determinant is to be expanded.
It is understood that (1.5) has to hold for {\it all\/} values of these
coefficients. Hence, one obtains $d-1\choose D-1$ equations -- clearly
generalising the Bateman equation (1.2) for one field -- whose general
covariance under arbitrary field redefinitions in $\phi\sp a$ is easily
established. Even though these equations form an over determined set, except
for $(D=1)$ or $(D=d-1)$ (the equations in the latter case were
shown$\sp {[1]}$
to be also universal), their space of solutions is non empty$\sp {[1]}$.
Functions $\phi\sp a(x_i)$ defined implicitly through the $D\ts$ constraints
$$x_iF\sp a_i(\phi\sp b)=c\sp a,\eqno(1.6)$$
where $F\sp a_i(\phi\sp b)$ are arbitrary functions of the fields
and $c\sp a$ arbitrary
constants, indeed always provide solutions to (1.5). In fact, (1.6)
defines$\sp {[4,5]}$ the general solution to the Bateman equation in two
dimensions.

In the present paper, the above results and conjecture are completed and
extended in different directions. On the one hand, finite hierarchies
with the characteristics i) -- v) are shown to exist even when ii) is
replaced by
\item{iib)} The proportionality factor mentioned in i) is essentially
{\it any\/} function of the first derivatives of the fields.

Thus, the function by which the equations of motion at a given level are
multiplied to define the Lagrangian at the next level may differ from
level to level, and need not be the first Lagrangian in the hierarchy.
As long as these multiplicative factors are functions of first derivatives of
the fields only, as is the first Lagrangian, the hierarchy will be
{\it finite} and terminate with universal equations of motion as described
in v), namely, up
to a factor, these equations are {\it independent\/} of the choice of
multiplicative factors at every level. In particular in cases 1) -- 3), the
universal equations are still given by (1.1) -- (1.3), provided the
successive multiplicative factors in cases 2) and 3) possess the same
properties as the first Lagrangian. In the following, the general
construction leading to (1.1)
will be referred to as the {\it generic hierarchy\/}, whereas the
construction leading to (1.2) will be called the {\it Bateman hierarchy\/}.

On the other hand, some principles for the construction of classical
topological field theories -- as defined above -- and of reparametrisation
invariant field theories -- without introducing a metric field -- are
presented. These constructions are generalisations of the map used in Ref.[2]
to establish the existence of the hierarchy leading to (1.3) by relating it
to the Bateman hierarchy. Namely, there exists a transformation, called here
the {\it C-map\/}, taking any field theory of $p\ts$ fields depending
on $q\ts$ variables into a classical topological field theory of $p\ts$
fields depending on $(p+q)$ variables, with equations of motion generally
covariant under arbitrary redefinitions of the $p\ts$ fields. Similarly,
there exists a
transformation, called the {\it R-map\/}, taking any field theory of $p\ts$
fields in $q\ts$ variables into a reparametrisation invariant field theory of
$(p+q)$ fields in $q\ts$ variables. Moreover, these maps also define
transformations between the equations of motion of the corresponding theories,
thus providing relations between their respective solutions. Note that the
C-map always produces a classical topological field theory with fewer fields
than the number of coordinates, whereas the R-map always produces a
reparametrisation invariant field theory with more fields than coordinates.

In particular, cases 1) -- 3) are transformed into one another under
applications of the C- and R-maps (these cases correspond to the situation
described above with $(p=1)$ and $(q=d\ts)$ arbitrary). Moreover, the C-map
may be used to transform the generic hierarchy into a new hierarchy with
(1.5) as its universal equations of motion. Under the R-map, one also obtains
universal equations of motion for a $(q-1)$-dimensional membrane
theory in a $(p+q)$-dimensional spacetime, thus in particular a
universal string theory in $(p+2)$ dimensions. If the suggestion that
these universal equations are integrable systems is correct, the R-map
would thus lead to new types of {\it integrable} membrane (and string)
theories of any dimension in a spacetime of arbitrary dimension, in
contradistinction to membrane theories based on the Nambu-Goto
action$\sp {[7,8]}$ whose equations of motion are linear in second derivatives
of the fields (these latter equations define an integrable system only in
the string case $(q=2)$).

As will become clearer later on, classical topological field theories and
reparametrisation invariant field theories -- of the type constructed in this
paper -- may be viewed as being dual to each other. This duality is realised
through the C- and R-maps. In some sense, the universal reparametrisation
invariant string and membrane equations of motion obtained here are
generalisations of the usual Nambu-Goto equations, and their universal dual
systems are classical topological field theories with fewer
fields than coordinates generalising the original two dimensional Bateman
equation$\sp {[4]}$. In fact, this idea of duality, now explicitly realised
through the C- and R-maps, was precisely one of the motivations in Ref.[1]
which led to the discovery of finite Euler hierarchies with universal
equations.

The paper is organised as follows. In sect.2, the relation between general
covariance of equations of motion under field redefinitions, reparametrisation
invariance and homogeneity properties of Lagrangians is discussed, extending
some of the considerations of Refs.[1,2]. Sects.3 and 4 present the
construction of C- and R-maps respectively, and discuss some of their basic
properties. In sect.5, a new proof for the generic Euler hierarchy is
presented, completing and generalising the main result of Ref.[1]. Using
the generic hierarchy, sect.6 shows how C- and R-maps can be applied to
obtain finite Euler hierarchies of classical topological field theories and of
reparametrisation invariant field theories, with the associated
universal equations of motion. In particular, this includes the hierarchies
leading to (1.2), (1.3) and (1.5). In sect.7, some general classes of
solutions to these universal equations are presented, taking advantage of the
C- and R-maps to relate solutions to different equations. Finally,
conclusions are presented in sect.8.  In an Appendix, definitions and
properties of generalised determinants and traces of matrices, as they appear
in the proof of the generic hierarchy, are collected for reference.

\vskip 20pt
\leftline{\bf 2.  Homogeneity, General Covariance and Reparametrisation
Invariance}
\vskip 20pt
In this section, we consider theories of $D\ts$ fields $\phi\sp a$ depending
on $d\ts$ coordinates $x_i$, with Lagrangian densities
${\cal L}(\phi\sp a_i,\phi\sp a_{ij})$ functions of first and second
derivatives of the fields only. Even though the results of this section
are valid for
Lagrangians depending on derivatives of arbitrary higher order, the
restriction to first and second derivatives only is sufficient for the
purposes of this paper. The equation of motion for the field $\phi\sp a$ is
thus given by
$${\cal E}_a{\cal L}[\phi\sp a]=0\ ,\eqno(2.1)$$
with the Euler operators ${\cal E}_a$ defined as
$${\cal E}_a{\cal L}[\phi\sp a]=
\pd_i\bigl[{\pd{\cal L}\over\pd\phi\sp a_i}(\phi\sp a_i,\phi\sp a_{ij})\bigr]-
\pd_i\pd_j\bigl[{\pd{\cal L}\over\pd\phi\sp a_{ij}}
(\phi\sp a_i,\phi\sp a_{ij})\bigr]\ .\eqno(2.2)$$
For most expressions in this paper, the arguments $[\phi\sp a]$ of an
equation of motion will not be made explicit, unless they are different
from the fields $\phi\sp a$.

First of all, consider Lagrangians with the following homogeneity property
$$
{\cal L}(R_i{}\sp j\phi\sp a_j,R_i{}\sp kR_j{}\sp l\phi\sp a_{kl}
+T\sp k_{ij}\phi\sp a_k) =
(\det R_i{}\sp j)\sp \alpha
{\cal L}(\phi\sp a_i,\phi\sp a_{ij})\ .\eqno(2.3)$$
Here, $R_i{}\sp j$ and $T\sp k_{ij}$ are arbitrary coefficients, and
$\alpha$ is the
weight of homogeneity of the Lagrangian. Differentiation of (2.3) with respect
to the parameters $R_i{}\sp j$ and $T\sp k_{ij}$ leads to the identities
$$
\eqalignno{\phi\sp a_j{\pd{\cal L}\over\pd\phi\sp a_i}
(\phi\sp a_i,\phi\sp a_{ij})
+\phi\sp a_{jk}\bigl[{\pd{\cal L}\over\pd\phi\sp a_{ik}}
(\phi\sp a_i,\phi\sp a_{ij})
&+{\pd{\cal L}\over\pd\phi\sp a_{ki}}(\phi\sp a_i,\phi\sp a_{ij})\bigr]=
\alpha\ \delta\sp i_j\ {\cal L}(\phi\sp a_i,\phi\sp a_{ij})\ ,&(2.4a)\cr
\phi\sp a_k{\pd{\cal L}\over\pd\phi\sp a_{ij}}(\phi\sp a_i,\phi\sp a_{ij})
&=0.&(2.4b)\cr}
$$
Note that even though $(\phi\sp a_{ij}=\phi\sp a_{ji})$, the dependence of the
Lagrangian on $\phi\sp a_{ij}$ is not assumed to be necessarily identical
to its dependence on $\phi\sp a_{ji}$, for fixed $(i\neq j)$.

Using these properties, a straightforward calculation shows that the equations
of motion obey the following relations
$$
\phi\sp a_i{\cal E}_a{\cal L}[\phi\sp a]=(\alpha-1)\ \pd_i{\cal L}
(\phi\sp a_i,\phi\sp a_{ij}).\eqno(2.5)
$$
Therefore, when $(\alpha=1)$ there are $d\ts$ identities among the $D\ts$
equations of motion, leaving only $(D-d\ts)$ independent equations of motion.
In fact, the case $(\alpha=1)$ corresponds to a reparametrisation invariant
action. Indeed, under a general reparametrisation of the coordinates of the
form
$$x_i\rightarrow y_i= y_i(x_j),\ \phi\sp a(x_i)\rightarrow
\widetilde\phi\sp a(y_i)=\phi\sp a(x_i),\eqno(2.6)$$
we have
$$\eqalign{{\pd\widetilde\phi\sp a\over\pd y_i}&=R_i{}\sp j
{\pd\phi\sp a\over\pd x_j}\ ,\cr
{\pd\sp 2\widetilde\phi\sp a\over\pd y_i\pd y_j}&=
R_i{}\sp kR_j{}\sp l{\pd\sp 2\phi\sp a\over\pd x_k\pd x_l}
+T_{ij}\sp k{\pd\phi\sp a\over\pd x_k}\ ,\cr}\eqno(2.7a)$$
with
$$R_i{}\sp j={\pd x_j\over\pd y_i}\ ,\quad\qquad T_{ij}\sp k
={\pd\sp 2x_k\over\pd y_i\pd y_j}\ .\eqno(2.7b)$$
Therefore, any Lagrangian obeying (2.3) scales under reparametrisations with
a factor $(\det R_i{}\sp j)\sp \alpha$. This factor cancels the Jacobian
$(\det R_i{}\sp j)\sp {-1}$ of the integration measure $\prod_i dx_i$ over the
coordinates precisely for $(\alpha=1)$. Hence in this case, the identities
$$\phi\sp a_i{\cal E}_a{\cal L}[\phi\sp a]=0\ ,\eqno(2.8)$$
are the Noether or Ward identities due to reparametrisation invariance,
leaving only $(D-d\ts)$ independent equations of motion. Incidentally, this
conclusion also shows that a {\it non trivial\/} reparametrisation invariant
field theory requires more fields than coordinates.

By analogy with (2.3), let us now consider Lagrangians with the following
homogeneity property
$${\cal L}(\phi\sp b_iR_b{}\sp a,\phi\sp b_{ij}R_b{}\sp a
   +\phi\sp b_iT\sp {ba}_j+\phi\sp b_jT\sp {ba}_i)=
(\det R_a{}\sp b)\sp \alpha {\cal L}(\phi\sp a_i,\phi\sp a_{ij})
\ .\eqno(2.9)$$
Here, $R_a{}\sp b$ and $T\sp {ab}_i$ are arbitrary coefficients,
and $\alpha$ is the
weight of homogeneity of the Lagrangian. Differentiation of (2.9) with
respect to the coefficients $R_a{}\sp b$ and $T\sp {ab}_i$ leads
to the identities
$$
\eqalignno{\phi\sp a_i{\pd{\cal L}\over\pd\phi\sp b_i}
(\phi\sp a_i,\phi\sp a_{ij})
+\phi\sp a_{ij}{\pd{\cal L}\over\pd\phi\sp b_{ij}}
(\phi\sp a_i,\phi\sp a_{ij})&=
\alpha\ \delta\sp a_b\ {\cal L}(\phi\sp a_i,\phi\sp a_{ij}),&(2.10a)\cr
\phi\sp a_j\bigl[{\pd{\cal L}\over\pd\phi\sp b_{ij}}
(\phi\sp a_i,\phi\sp a_{ij}) + {\pd{\cal L}\over\pd\phi\sp b_{ji}}
(\phi\sp a_i,\phi\sp a_{ij})\bigr]&
=0.&(2.10b)\cr}
$$

On the other hand, differentiation of (2.9) with respect
to $\phi\sp a_{ij}$ and
$\phi\sp a_i$ leads to the further homogeneity properties:
$$
\eqalignno{{\pd{\cal L}\over\pd\phi\sp a_{ij}}(\phi\sp b_iR_b{}\sp a,
\phi\sp b_{ij}R_b{}\sp a+\phi\sp b_iT\sp {ba}_j+\phi\sp b_jT\sp {ba}_i)&=
(\det R_a{}\sp b)\sp \alpha(R\sp {-1})_a{}\sp b{\pd{\cal L}
\over\pd\phi\sp b_{ij}}
(\phi\sp a_i,\phi\sp a_{ij}),&(2.11a)\cr
{\pd{\cal L}\over\pd\phi\sp a_i}
(\phi\sp b_iR_b{}\sp a,\phi\sp b_{ij}R_b{}\sp a
+\phi\sp b_iT\sp {ba}_j+\phi\sp b_jT\sp {ba}_i)&=
(\det R_a{}\sp b)\sp \alpha(R\sp {-1})_a{}\sp b
{\pd{\cal L}\over\pd\phi\sp b_i}(\phi\sp a_i,\phi\sp a_{ij})-\cr
-(\det R_a{}\sp b)\sp \alpha(R\sp {-1})_a{}\sp bT\sp {bc}_j
(R\sp {-1})_c{}\sp d
&\bigl[ {\pd{\cal L}\over\pd\phi\sp d_{ij}}(\phi\sp a_i,\phi\sp a_{ij})
+ {\pd{\cal L}\over\pd\phi\sp d_{ji}}
(\phi\sp a_i,\phi\sp a_{ij})\bigr].&(2.11b)\cr}
$$

Given these properties, consider now the transformation of the equations of
motion (2.2) under arbitrary field redefinitions
$$\varphi\sp a=F\sp a(\phi\sp b)\ .\eqno(2.12)$$
We then have
$$\varphi\sp a_i=\phi\sp b_iR_b{}\sp a\ ,
\qquad{\varphi\sp a_{ij}=\phi\sp b_{ij}R_b{}\sp a+\phi\sp b_iT\sp {ba}_j
 +\phi\sp b_jT\sp {ba}_i},\eqno(2.13a)$$
where
$$R_a{}\sp b={\pd F\sp b\over\pd\phi\sp a}\ ,
\quad\qquad{T\sp {ba}_i={1\over2}\phi\sp c_i
{\pd\sp 2F\sp a\over\pd\phi\sp b\pd\phi\sp c}}\ .
\eqno(2.13b)$$
Using the homogeneity properties (2.11) and the identities (2.10), a
straightforward calculation then shows that the equations of motion
${\cal E}_a{\cal L}[\varphi\sp a]$ for the transformed fields
$\varphi\sp a$ are
given in terms of those for the fields $\phi\sp a$ by
$${\cal E}_a{\cal L}[\varphi\sp a]=
(\det R_a{}\sp b)\sp \alpha (R\sp {-1})_a{}\sp b {\cal E}_b
{\cal L}[\phi\sp a]
+(R\sp {-1})_a{}\sp b{\pd(\det R_a{}\sp b)\sp \alpha
\over\pd\phi\sp b}(\alpha-1)
{\cal L}(\phi\sp a_i,\phi\sp a_{ij})\ .\eqno(2.14)$$

Therefore, whenever the Lagrangian possesses the homogeneity property (2.9)
with a weight $(\alpha=0)$ or $(\alpha=1)$, the equations of motion for the
fields $\phi\sp a$ transform covariantly among themselves under arbitrary
field redefinitions (2.12). This result is the complete generalisation of
theorems in Ref.[1] establishing the same property in the case of a Lagrangian
${\cal L}(\phi\sp a_i)$ function only of first derivatives of the fields. In
particular in the case of one field ($D=1$), the equation of motion (2.2) is
{\it invariant} when $(\alpha=1)$.

To conclude, we have thus shown that any Lagrangian
${\cal L}(\phi\sp a_i,\phi\sp a_{ij})$ homogeneous in the sense of (2.9)
with a weight $(\alpha=1)$ or $(\alpha=0)$ defines a classical topological
field theory. In fact, the case $(\alpha=1)$ plays a distinguished
r$\hat{\rm o}$le, as will become clear in the next section when
discussing the C-map.

\vskip 20pt
\leftline{\bf 3. The C-Map}
\vskip 20pt
To define the C-map, let us consider an arbitrary field theory Lagrangian
${\cal L}(y\sp a_\alpha,y\sp a_{\alpha\beta})$ dependent on the first and
second
derivatives of $p\ts$ fields $y\sp a(x_\alpha)$ functions of $q\ts$
coordinates
$x_\alpha$. The associated action is thus
$$S[y\sp a]=\int{\prod_\alpha dx_\alpha \ {\cal L}
({\pd y\sp a\over\pd x_\alpha},
{\pd\sp 2y\sp a\over\pd x_\alpha\pd x_\beta})}\ ,\eqno(3.1)$$
whose variation under arbitrary field variations $\delta y\sp a$ is given,
up to surface terms, by
$$\delta S[y\sp a]=-\int{\prod_\alpha dx_\alpha \ \delta y\sp a
\ {\cal E}_{y\sp a}{\cal L}}\ .\eqno(3.2)$$
The Euler operators are
$${\cal E}_{y\sp a}{\cal L}=
\pd_\alpha{\pd{\cal L}\over\pd y\sp a_\alpha}
-\pd_\alpha\pd_\beta{\pd{\cal L}\over\pd y\sp a_{\alpha\beta}}\ .\eqno(3.3)$$

Let us now introduce $p\ts$ additional coordinates $\phi\sp a$, and extend the
coordinate dependence of the fields $y\sp a$ to also include a dependence on
these new variables, {\it i.e.} $y\sp a(x_\alpha,\phi\sp b)$. Nevertheless,
the corresponding field theory is still described by the original Lagrangian
${\cal L}(y\sp a_\alpha,y\sp a_{\alpha\beta})$, with the action now given by
$$S[y\sp a]=\int{\prod_\alpha dx_\alpha \prod_a d\phi\sp a \ {\cal L}
({\pd y\sp a\over\pd x_\alpha},{\pd\sp 2y\sp a\over\pd x_\alpha\pd x_\beta})}
\ .\eqno(3.4)$$
In particular, field variations still lead to
$$\delta S[y\sp a]=-\int{\prod_\alpha dx_\alpha \prod_a d\phi\sp a
\ \delta y\sp a\ {\cal E}_{y\sp a}{\cal L}}\ ,\eqno(3.5)$$
where the Euler operators ${\cal E}_{y\sp a}$ have the {\it same\/}
definition as in (3.3), since the Lagrangian is {\it independent\/} of
any derivatives of the fields $y\sp a$ with respect to the new
variables $\phi\sp a$. In other words, these variables are irrelevant
as far as the dynamical evolution of the fields $y\sp a$ is concerned;
the equations of motion for both actions (3.1) and (3.4) are identical.

The intention is to invert the $\phi\sp a$ dependence of the
fields $y\sp a$, and to
consider $\phi\sp a$ as functions of $(x_i=(x_\alpha,y\sp a))$, {\it i.e.}
$\phi\sp a(x_\alpha,y\sp a)$. This requires that the matrix of derivatives
$({\pd y\sp a}/{\pd\phi\sp b})$ be non singular, or equivalently that we have
$$\det M_a{}\sp b\neq 0\ ,\eqno(3.6)$$
with $M_a{}\sp b$ being the inverse matrix of derivatives
$({\pd y\sp a}/{\pd\phi\sp b})$ keeping $x_\alpha$ fixed, namely
$$M_a{}\sp b={\pd\phi\sp b\over\pd y\sp a}\ .\eqno(3.7)$$
The $\phi\sp a$ dependence of the fields $y\sp a$ may then be inverted,
leading to
configurations $\phi\sp a(x_i)$ for $(D=p)$ fields $\phi\sp a$ functions of
$(d=p+q)$ variables $(x_i=(x_\alpha,y\sp a))$. By direct substitution in the
Lagrangian for the original field theory and its action (3.4), one then
obtains a new Lagrangian field theory for the fields
$\phi\sp a(x_\alpha,y\sp b)$. As is
shown hereafter, the resulting theory is actually a classical topological
field
theory. A fortiori, this is to be expected. Indeed, as was pointed out above,
the $\phi\sp a$ dependence in the original theory is irrelevant, so that the
equations of motion for the fields $\phi\sp a$ in the new theory should be
independent of the parametrisation used for these fields, namely their
equations of motion should be generally covariant under arbitrary field
transformations.

The series of operations described above define the C-map, which
thus takes any field theory of $p\ts$ fields in $q\ts$ dimensions into a
classical topological field theory of $(D=p)$ fields in $(d=p+q)$ dimensions.
Note that the discussion is developed here for Lagrangians depending on first
and second derivatives of fields only. However, it should be clear that the
same considerations and conclusions apply in the case of Lagrangians
depending on derivatives of arbitrary high order.

{}From the inversion $\phi\sp a(x_\alpha,y\sp b)$ defining the C-map, it is
straightforward to derive the expressions that substitute for the derivatives
of the original fields $y\sp a$. Namely, the derivatives
$({\pd y\sp a}/{\pd x_\alpha})$ are substituted by the quantities
$$Y\sp a_\alpha=-{\pd\phi\sp b\over\pd x_\alpha}
(M\sp {-1})_b{}\sp a,\eqno(3.8)$$
whereas the derivatives $({\pd\sp 2y\sp a}/{(\pd x_\alpha\pd x_\beta)})$ are
transformed into the quantities
$$\eqalign{
Y\sp a_{\alpha\beta}=-\bigl[{\pd\sp 2\phi\sp b
\over\pd x_\alpha\pd x_\beta}&-
{\pd\phi\sp c\over\pd x_\alpha}(M\sp {-1})_c{}\sp d
{\pd\sp 2\phi\sp b\over\pd y\sp d\pd x_\beta}
-{\pd\phi\sp c\over\pd x_\beta}(M\sp {-1})_c{}\sp d
{\pd\sp 2\phi\sp b\over\pd y\sp d\pd x_\alpha}+\cr
&+{\pd\phi\sp c\over\pd x_\alpha}(M\sp {-1})_c{}\sp e
{\pd\phi\sp d\over\pd x_\beta}(M\sp {-1})_d{}\sp f
{\pd\sp 2\phi\sp b\over\pd y\sp e\pd y\sp f}\bigr]
(M\sp {-1})_b{}\sp a\ .\cr}\eqno(3.9)$$

Given these definitions, the Lagrangian for the transformed theory with
fields $\phi\sp a(x_i)$ and coordinates $(x_i=(x_\alpha,y\sp a))$ is
simply obtained in terms of the original Lagrangian
${\cal L}(y\sp a_\alpha,y\sp a_{\alpha\beta})$ as
$${\cal L}_{C}(\phi\sp a_i,\phi\sp a_{ij})=
(\det M_a{}\sp b)\ {\cal L}(Y\sp a_\alpha,Y\sp a_{\alpha\beta})
\ .\eqno(3.10)$$
With this identification, it is quite clear that the Lagrangian
${\cal L}_{C}(\phi\sp a_i,\phi\sp a_{ij})$ possesses the homogeneity
property (2.9) with a weight $(\alpha=1)$, thus establishing at once that
the C-map has indeed produced a classical topological field theory.

By consideration of the variations of the actions for both field theories
under variations of the fields $y\sp a$ and $\phi\sp a$, one also obtains
the relation between the respective field equations of motion as
$${\cal E}_{\phi\sp a}{\cal L}_{C}=
-(\det M_a{}\sp b)(M\sp {-1})_a{}\sp b\ {\cal E}_{y\sp b}{\cal L}
\ ,\eqno(3.11)$$
where it is of course understood that the appropriate substitutions, such as
those in (3.8) and (3.9), have to be applied on the right-hand side. Hence,
solutions of either set of equations for which the non degeneracy condition
(3.6) is satisfied are in one-to-one correspondence through the C-map. Note
that the relations (3.11) also confirm the general covariance under field
transformations of the equations for the fields $\phi\sp a$. It is the
prefactor $((\det M_a{}\sp b)(M\sp {-1})_a{}\sp b)$ in (3.11) which
governs the transformation of
these equations -- and this prefactor indeed transforms covariantly
under field redefinitions -- since the equations
$({\cal E}_{y\sp a}{\cal L})$ are independent
of the parametrisation used for $\phi\sp a$.

Applying the C-map once more on the theory described by
${\cal L}_{C}(\phi\sp a_i,\phi\sp a_{ij})$ does not produce a new classical
topological field theory. Indeed, if the original field theory with Lagrangian
${\cal L}(y\sp a_\alpha,y\sp a_{\alpha\beta})$ is itself a classical
topological
field theory in the sense of (2.9) with a weight $(\alpha=1)$, from that
property and the identification (3.10) it is clear that the net effect of
the C-map is to change the sign of the action by $(-1)\sp p$ and to substitute
the fields $y\sp a$ by the fields $\phi\sp a$. In other words, no
new field theory
is produced, in agreement with the fact that the inversion defining the
C-map is essentially a redefinition of the fields $y\sp a$, a transformation
which does not affect the equations of motion for a classical topological
field theory.

It is also possible to invert the C-map. Namely, given any
classical topological field theory of $(D=p)$ fields dependent on $(d=p+q)$
variables, it is possible to define a new field theory of $p\ts$ fields
dependent on $q\ts$ variables whose image under the C-map reproduces the
initial classical topological field theory. Specifically, consider a
Lagrangian ${\cal L}_{C}(\phi\sp a_i,\phi\sp a_{ij})$ with $(D=p)$ fields
$\phi\sp a(x_i)$ of $(d=p+q)$ coordinates $(x_i=(x_\alpha,y\sp a))$
(there are thus
$p\ts$ coordinates $y\sp a$ and $q\ts$ coordinates $x_\alpha$). This
Lagrangian
is assumed to obey the homogeneity property (2.9) with weight $(\alpha=1)$.

As before, in order to invert the $y\sp a$ dependence, consider field
configurations such that
$$\det N_a{}\sp b\neq 0\ ,\eqno(3.12)$$
where the matrix $N_a{}\sp b$ is defined to be
$$N_a{}\sp b={\pd y\sp b\over\pd\phi\sp a}\ .\eqno(3.13)$$
The $y\sp a$ dependence of the fields $\phi\sp a$ may then be inverted,
leading to
the fields $y\sp a(x_\alpha,\phi\sp b)$. By direct substitution in the
Lagrangian
${\cal L}_{C}$, one then obtains an action principle for these fields.
However, due to the classical general covariance under field redefinitions of
the system described by ${\cal L}_{C}$, one expects that any dependence on
$\phi\sp a$ actually decouples in the equations of motion for the fields
$y\sp a$.
As is shown hereafter, such a decoupling indeed occurs. Hence, the
inversion of
the C-map as described above, takes any classical topological field theory of
$p\ts$ fields in $(p+q)$ dimensions into a field theory of $p\ts$ fields in
$q\ts$ dimensions. Applying then the C-map on this latter field theory, one
recovers the initial classical topological field theory described by
${\cal L}_{C}$. Here again, the same conclusions would apply to Lagrangians
depending on field derivatives of higher order.

{}From the inversion $y\sp a(x_\alpha,\phi\sp a)$ defining the inverse
C-map, it is
straightforward to derive the expressions that substitute for the derivatives
of the original fields $\phi\sp a$. The Lagrangian for the fields
$y\sp a$ is then
given by the function
$$(\det N_a{}\sp b)\ {\cal L}_{C}(\Phi\sp b_iR_b{}\sp a,
\Phi\sp b_{ij}R_b{}\sp a+\Phi\sp b_iT\sp {ba}_j+\Phi\sp b_j
T\sp {ba}_i),\eqno(3.14)$$
where the quantities $\Phi\sp a_i$ and $\Phi\sp a_{ij}$ are defined as
$$\eqalign{\Phi\sp c_\alpha=-{\pd y\sp c\over\pd x_\alpha}\ ,&
\qquad\Phi\sp c_a=\delta_a\sp c\ ,\cr
\Phi\sp c_{\alpha\beta}=-{\pd\sp 2y\sp c\over\pd x_\alpha\pd x_\beta}\ ,\quad
\Phi\sp c_{\alpha b}&=0\ ,\quad\Phi\sp c_{a\beta}=0\ ,\quad\Phi\sp c_{ab}=0
\ ,\cr}\eqno(3.15)$$
and the coefficients $R_a{}\sp b$ and $T\sp {ab}_i$ by
$$\eqalign{R_a{}\sp b=&\ (N\sp {-1})_a{}\sp b,\cr
T\sp {bc}_\alpha=&\ (N\sp {-1})_b{}\sp d\bigl[{1\over2}
{\pd y\sp e\over\pd x_\alpha}
(N\sp {-1})_e{}\sp f{\pd\sp 2y\sp a\over\pd\phi\sp d\pd\phi\sp f}
-{\pd\sp 2y\sp a\over\pd x_\alpha\pd\phi\sp d}\bigr](N\sp {-1})_a{}\sp c,\cr
T\sp {bc}_a=&-{1\over2}(N\sp {-1})_a{}\sp e(N\sp {-1})_b{}\sp f
{\pd\sp 2y\sp d\over\pd\phi\sp e\pd\phi\sp f}(N\sp {-1})_d{}\sp c\ .}
\eqno(3.16)$$
However, owing to the homogeneity properties of the Lagrangian
${\cal L}_{C}$, (3.14) simplifies considerably, giving finally as Lagrangian
for the fields $y\sp a$ the function
$${\cal L}({\pd y\sp a\over\pd x_\alpha},
{\pd\sp 2y\sp a\over\pd x_\alpha\pd x_\beta})=
{\cal L}_{C}(\Phi\sp a_i,\Phi\sp a_{ij})\ .\eqno(3.17)$$
Since the quantities $\Phi\sp a_i$ and $\Phi\sp a_{ij}$ only involve
derivatives of
the fields $y\sp a$ with respect to the variables $x_\alpha$, the $\phi\sp a$
coordinates effectively decouple from the system, and one indeed
obtains a theory of $p\ts$ fields $y\sp a$ dependent on $q\ts$
coordinates $x_\alpha$. Namely, suppressing from the action the
integration over the variables $\phi\sp a$ does not modify the equations
of motion for the fields $y\sp a$, as was already pointed out after (3.4)
and (3.5) when considering the C-map.

Obviously, there is also a direct relation between the equations of motion for
the two field theories related through the inverse C-map. This relation is
essentially the inverse of (3.11), and reads
$${\cal E}_{y\sp a}{\cal L}=
-(\det N_a{}\sp b)(N\sp {-1})_a{}\sp b\ {\cal E}_{\phi\sp b}{\cal L}_{C}\ .
\eqno(3.18)$$

Finally, it is easy to verify that applying the C-map on the field
theory described by ${\cal L}(y\sp a_\alpha,y\sp a_{\alpha\beta})$
in (3.17) gives
back the initial classical topological field theory described by
${\cal L}_{C}(\phi\sp a_i,\phi\sp a_{ij})$. In particular, note that this
conclusion also implies that {\it any\/} classical topological field theory
of $p\ts$ fields depending on $(p+q)$ coordinates and obeying the homogeneity
property (2.9) with weight $(\alpha=1)$, may {\it always\/} be viewed as
resulting from the C-map acting on some specific but otherwise arbitrary
field theory Lagrangian of $p\ts$ fields depending on $q\ts$ coordinates. The
C-map puts these two classes of fields theories in one-to-one correspondence.

\vskip 20pt
\leftline{\bf 4. The R-Map}
\vskip 20pt
To define the R-map, consider now an arbitrary field theory Lagrangian
${\cal L}({\pd\phi\sp \mu\over\pd z_i},
{\pd\sp 2\phi\sp \mu\over\pd z_i\pd z_j})$
dependent on the first and second derivatives of $p\ts$ fields
$\phi\sp \mu(z_i)$
functions of $q\ts$ coordinates $z_i$. The associated action is thus
$$S[\phi\sp \mu]=\int{\prod_i dz_i \ {\cal L}
({\pd\phi\sp \mu\over\pd z_i},{\pd\sp 2\phi\sp \mu\over\pd z_i\pd z_j})}\ .
\eqno(4.1)$$
As always, variations of this action under field variations
$\delta\phi\sp \mu$ are given, up to surface terms, by
$$\delta S[\phi\sp \mu]=-\int{\prod_i dz_i\ \delta\phi\sp \mu\
{\cal E}_{\phi\sp \mu}{\cal L}}\ ,\eqno(4.2)$$
with the Euler operators
$${\cal E}_{\phi\sp \mu}={\pd\over\pd z_i}
\bigl[{\pd{\cal L}\over\pd({\pd\phi\sp \mu\over\pd z_i})}\bigr]-
{\pd\over\pd z_i}{\pd\over\pd z_j}
\bigl[{\pd{\cal L}\over\pd
({\pd\sp 2\phi\sp \mu\over\pd z_i\pd z_j})}\bigr]\ .
\eqno(4.3)$$

Let us now introduce $q\ts$ arbitrary functions $x_i(z_j)$, and extend the
field content of the theory to also include these extra degrees of freedom,
while still keeping the same Lagrangian and action as in (4.1). Obviously,
the equations of motion for $\phi\sp \mu$ remain as before, whereas those
for the new fields $x_i$ are trivially satisfied since the Lagrangian is
independent of these degrees of freedom.

Nevertheless, by inverting the $z_i$ dependence of the theory, one obtains a
new field theory with reparametrisation invariance. For this purpose, consider
field configurations such that
$${\det M_i{}\sp j}\neq 0\ ,\eqno(4.4)$$
where the matrix $M_i{}\sp j$ is defined to be
$$M_i{}\sp j={\pd z_j\over\pd x_i}\ .\eqno(4.5)$$
The $z_i$ dependence of the fields $x_i$ may then be inverted, leading to a
field theory of $(D=p+q)$ fields $(\phi\sp a=(\phi\sp \mu,z_i))$ functions of
$(d=q)$ variables $x_i$, with an action obtained from (4.2) by direct
substitution. As shown hereafter, the resulting field theory is
reparametrisation invariant in the coordinates $x_i$. A priori, this fact is
to be expected since, on the one hand, the choice for the functions
$x_i(z_j)$ is totally arbitrary, and on the other hand, their equations of
motion are trivial, so that the transformed theory should indeed be
independent of the choice of parametrisation in $x_i$.
{}From a geometric point of view, the functions $\phi\sp \mu(z_i)$ define a
$q\ts$-dimensional subspace of the $(p+q)$-dimensional space spanned by the
coordinates $(\phi\sp a=(\phi\sp \mu,z_i))$. Introducing the fields $x_i(z_j)$
amounts to introducing an arbitrary intrinsic parametrisation of this
subspace, without affecting its topological and geometrical properties as
an embedded space. In other words, we are simply dealing with parametrised
$(q-1)$-dimensional membrane theories in $(p+q)$ dimensions.

The series of operations described above define the R-map, which thus takes
any field theory of $p\ts$ fields in $q\ts$ dimensions into a
reparametrisation invariant field theory of $(D=p+q)$ fields in $(d=q)$
dimensions. Again,
even though the discussion is developed for Lagrangians depending on first
and second derivatives of the fields only, the same conclusions apply in the
case of Lagrangians depending on derivatives of arbitrary high order.

{}From the inversion $\phi\sp a(x_i)$ defining the R-map, it is easy to derive
the expressions that substitute for the derivatives of the original fields
$\phi\sp \mu$ with respect to the variables $z_i$. Namely, the derivatives
$({\pd\phi\sp \mu}/{\pd z_i})$ are replaced by
$$Y\sp \mu_i=(M\sp {-1})_i{}\sp j{\pd\phi\sp \mu\over\pd x_j}\ ,\eqno(4.6)$$
whereas the derivatives $({\pd\sp 2\phi\sp \mu}/{(\pd z_i\pd z_j)})$ are
substituted by the quantities
$$Y\sp \mu_{ij}=(M\sp {-1})_i{}\sp k(M\sp {-1})_j{}\sp l\bigl[
{\pd\sp 2\phi\sp \mu\over\pd x_k\pd x_l}-
{\pd z_m\over\pd x_k\pd x_l}(M\sp {-1})_m{}\sp n
\ {\pd\phi\sp \mu\over\pd x_n}\bigr]\ .\eqno(4.7)$$

Given these definitions, the Lagrangian for the fields
$(\phi\sp a(x_j)=(\phi\sp \mu(x_j),z_i(x_j)))$ and the coordinates $x_i$
is simply obtained in terms of the original Lagrangian ${\cal L}$ in (4.1) as
$${\cal L}_R({\pd\phi\sp a\over\pd x_i},
{\pd\sp 2\phi\sp a\over\pd x_i\pd x_j})=
(\det M_i{}\sp j)\ {\cal L}(Y_i\sp \mu,Y_{ij}\sp \mu)\ .\eqno(4.8)$$
With this identification, it is straighforward to verify that the Lagrangian
${\cal L}_R$ obeys the homogeneity property (2.3) with a weight $(\alpha=1)$,
thereby establishing at once that the R-map has indeed produced a
reparametrisation invariant field theory.

The equations of motion of both theories are also in correspondence. However,
we know from (2.8) that for the reparametrisation invariant theory there are
only $p\ts$ independent equations, the remaining $q\ts$ equations being
obtained from the Ward identities. Clearly, the independent equations are
those for the $p\ts$ fields $\phi\sp \mu$, which for both theories are
related as
$${\cal E}_{\phi\sp \mu}{\cal L}_R=(\det M_i{}\sp j)\ {\cal E}_{\phi\sp \mu}
{\cal L}\ ,\eqno(4.9)$$
with of course the appropriate substitutions being implemented on the
right-hand side. The remaining equations of motion for the $q\ts$ fields
$z_i(x_j)$ are simply obtained from the Ward identities, giving
$$\eqalign{{\cal E}_{z_i}{\cal L}_R=&-(M\sp {-1})_i{}\sp j
{\pd\phi\sp \mu\over\pd x_j}
{\cal E}_{\phi\sp \mu}{\cal L}_R\cr
=&-(\det M_i{}\sp j)(M\sp {-1})_i{}\sp j{\pd\phi\sp \mu\over\pd x_j}
{\cal E}_{\phi\sp \mu}{\cal L}\ .\cr}\eqno(4.10)$$
These expressions also make it clear that under reparametrisations of the
coordinates $x_i$, the equations of motion for the theory obtained through the
R-map transform among themselves, in agreement with the reparametrisation
invariance of that theory.

Applying the R-map once more on the theory described by ${\cal L}_R$ does not
produce a new reparametrisation invariant theory. Indeed, if the original
theory is already reparametrisation invariant, namely satisfies (2.3) with
weight $(\alpha=1)$, it is clear from (4.6) -- (4.8) that the net effect of
the R-map is to substitute the $z_i$ dependence of the fields
$\phi\sp \mu$ by a
$x_i$ dependence, while the fields $z_i(x_j)$ that would otherwise arise from
the R-map actually do not appear in the transformed Lagrangian. Of course,
this is due to reparametrisation invariance, since the inversion defining
the R-map is simply a reparametrisation in the coordinates $z_i$.
The interest of the R-map is that it produces in the general case a
reparametrisation invariant field theory from one which is not.

Conversely, it is also possible to invert the R-map. Namely, given any
reparametrisation invariant theory of $(D=p+q)$ fields
$(\phi\sp a=(\phi\sp \mu,z_i))$ dependent on $(d=q)$ coordinates $x_i$ with
Lagrangian
${\cal L}_R({\pd\phi\sp a\over\pd x_i},
{\pd\sp 2\phi\sp a\over\pd x_i\pd x_j})$ (there
are thus $p\ts$ fields $\phi\sp \mu$ and $q\ts$ fields $z_i$,
and the Lagrangian
is assumed to obey (2.3) with $(\alpha=1)$), there is a map
producing a theory of $p\ts$ fields in $q\ts$ dimensions whose image under
the R-map is the initial theory. Clearly, the inverse R-map is obtained by
inverting the $x_i$ dependence of the fields $z_i$. Let us thus consider
field configurations such that
$$\det N_i{}\sp j\neq 0\ ,\eqno(4.11)$$
with the matrix $N_i{}\sp j$ being
$$N_i{}\sp j={\pd x_j\over\pd z_i}\ .\eqno(4.12)$$
The $x_i$ dependence is then invertible, leading to the fields
$\phi\sp \mu(z_j)$
and $x_i(z_j)$ with an action obtained from ${\cal L}_R$ by direct
substitution. Clearly, due to reparametrisation invariance of the latter
Lagrangian, one expects that the fields $x_i$ decouple from the system,
leaving a theory of $p\ts$ fields in $q\ts$ dimensions. Acting with the R-map
on the latter field theory reproduces again the initial reparametrisation
invariant one.

Specifically, the Lagrangian obtained from the inverse R-map is simply
$$(\det N_i{}\sp j)\ {\cal L}_R(R_i{}\sp j\Psi\sp a_j,
R_i{}\sp kR_j{}\sp l\Psi\sp a_{kl}+T\sp k_{ij}\Psi\sp a_k)\ ,\eqno(4.13)$$
where we defined
$$\eqalign{\Psi\sp \nu_i={\pd\phi\sp \nu\over\pd z_i}\ ,
&\quad\Psi\sp j_i=\delta_i\sp j\ ,\cr
\Psi\sp \rho_{ij}={\pd\sp 2\phi\sp \rho\over\pd z_i\pd z_j}\ ,
&\quad\Psi\sp k_{ij}=0\ ,\cr}\eqno(4.14)$$
and
$$\eqalign{R_i{}\sp j=&\ (N\sp {-1})_i{}\sp j\ ,\cr
T\sp k_{ij}=&-(N\sp {-1})_i{}\sp l(N\sp {-1})_j{}\sp m
{\pd\sp 2 x_n\over\pd z_l\pd z_m}
(N\sp {-1})_n{}\sp k\ .\cr}\eqno(4.15)$$
However, owing to the homogeneity properties of the Lagrangian ${\cal L}_R$,
(4.13) simplifies leaving for the transformed Lagrangian
$${\cal L}({\pd\phi\sp \mu\over\pd z_i},
{\pd\sp 2\phi\sp \mu\over\pd z_i\pd z_j})=
{\cal L}_R(\Psi\sp a_i,\Psi\sp a_{ij})\ .\eqno(4.16)$$
Clearly, since the quantities $\Psi\sp a_i$ and $\Psi\sp a_{ij}$ involve
derivatives
only of the fields $\phi\sp \mu(z_j)$, the fields $z_i(x_j)$ have indeed
decoupled, leaving a theory of $p\ts$ fields in $q\ts$ dimensions with the
Lagrangian (4.16). Obviously, the equations of motion for both theories are
also in correspondence, the relation being
$${\cal E}_{\phi\sp \mu}{\cal L}=
(\det N_i{}\sp j)\ {\cal E}_{\phi\sp \mu}{\cal L}_R\ .\eqno(4.17)$$
Finally, it is straightforward to verify that the R-map acting on the theory
defined by (4.16) gives back the original field theory described by
${\cal L}_R$. In particular, this result implies that {\it any}
reparametrisation invariant theory of $(p+q)$ fields in $q\ts$ dimensions,
with a Lagrangian obeying the homogeneity property (2.3) with $(\alpha=1)$,
may {\it always} be viewed as resulting from the R-map acting on some
specific but otherwise arbitrary theory of $p\ts$ fields in $q\ts$ dimensions.
The R-map puts these two classes of theories in one-to-one correspondence.

Having introduced the C- and R-maps and the associated inverse
transformations, one may wonder what happens when these maps are applied in
succession. We have already seen that applying twice the C-map or the R-map
only reproduces the result of the C- or the R-map applied once.
This only leaves the successive application of the two maps alternatively.
For convenience, given a Lagrangian ${\cal L}$ (with its field content
implicit), let us symbolically denote the results of the C- and R-maps as
$${\cal L}_{C}={\cal O}_{C}{\cal L}\ ,\qquad
{\cal L}_R={\cal O}_R{\cal L}\ .\eqno(4.18)$$
Thus, one may still act on ${\cal L}_{C}$ with the R-map and on ${\cal L}_R$
with the C-map. However, due to the respective properties -- under field
redefinitions or reparametrisations -- of these Lagrangians, it is possible
to show that the net result of these transformations is to lead essentially to
the following identifications:
$${\cal O}_R{\cal L}_{C}={\cal L}_R\ ,\qquad
{\cal O}_{C}{\cal L}_R={\cal L}_{C}\ .\eqno(4.19)$$
Namely, in either case the resulting system possesses one of the two properties
for all of the associated variables, and the other property for some subset
only of the other set of associated variables. Taking advantage of both
properties under field redefinitions and reparametrisations, it becomes
possible to set equal those fields and those coordinates for which both
properties are operative, so that these field degrees of freedom effectively
decouple from the system. This decoupling then only leaves the field
theories whose content is expressed through (4.19). Of course, the counting
of the number of equations of motion and Ward identities in all cases agrees
with this conclusion, as may easily be checked.

We have therefore obtained the following situation. Given an arbitrary
Lagrangian ${\cal L}$, the C- and R-maps produce only two other systems,
namely those specified by the Lagrangians ${\cal L}_{C}$ and ${\cal L}_R$,
the former being a classical topological field theory and the latter a
reparametrisation invariant one. These three Lagrangians belong to a
triangular set of fields theories which is closed under the action of the
C- and R-maps and their inverse maps. The C-map acting on all three
Lagrangians produces ${\cal L}_{C}$. The R-map acting on all three
Lagrangians produces ${\cal L}_R$. And finally, the Lagrangian ${\cal L}$
is obtained from ${\cal L}_{C}$ and from ${\cal L}_R$ through the action
of the inverse C- and R-maps respectively. This triangular relationship is the
precise expression of the duality existing between classical topological field
theories and reparametrisation invariant field theories mentioned in the
introduction.

\vskip 20pt
\leftline{\bf 5. The Generic Hierarchy}
\vskip 20 pt
Consider now a collection $F_n(\phi_i)\ (n=1,2,\ldots)$ of arbitrary functions
of the first derivatives of a field $\phi(x_i)$ dependent on $d\ts$
coordinates $x_i$. The associated Euler hierarchy of Lagrangians is
then defined recursively by
$${\cal L}_n=F_n\ W_{n-1},\qquad W_0=1\ ,\eqno(5.1)$$
where the $W_n\ (n=1,2,\ldots)$ are the equations of motion
$$W_n={\cal E}{\cal L}_n\ ,\eqno(5.2)$$
and ${\cal E}$ is the Euler operator defined in (2.2).

This describes the construction of the generic finite Euler hierarchy whose
properties where discussed in the introduction. This section provides proofs
for those properties. The fundamental result is the following identity
$$W_n={n!}\ Tr\sp {(n)}(A\sp {(1)},\cdots,A\sp {(n)})\ ,\eqno(5.3)$$
where
$$A\sp {(p)}_{ij}=M\sp {(p)}_{ik}\phi_{kj}\ ,\qquad M\sp {(p)}_{ij}=
{\pd\sp 2 F_p\over\pd\phi_i\pd\phi_j}\ ,\quad p=1,2,\cdots,n\ ,\eqno(5.4)$$
and the function $Tr\sp {(n)}$ in (5.3) is the generalised determinant
defined in (A.5) in the Appendix. In fact, explicit expressions for $W_n$ are
$$\eqalign{W_n=&\ T\sp {(n)}_{i_1\cdots i_n;j_1\cdots j_n}
A\sp {(1)}_{i_1 j_1}\cdots A\sp {(n)}_{i_n j_n}\cr
=&\ {1\over(d-n)!}\ \epsilon_{i_1\cdots i_d}\ \epsilon_{j_1\cdots j_d}
A\sp {(1)}_{i_1 j_1}\cdots A\sp {(n)}_{i_n j_n}
\ \delta_{i_{n+1}j_{n+1}}\cdots\delta_{i_d j_d}\cr
=&\ {1\over(d-n)!}\ \epsilon_{i_1\cdots i_d}\ \epsilon_{j_1\cdots j_d}
M\sp {(1)}_{i_1 k_1}\cdots M\sp {(n)}_{i_n k_n}
\ \phi_{k_1 j_1}\cdots\phi_{k_n j_n}
\ \delta_{i_{n+1}j_{n+1}}\cdots\delta_{i_d j_d}\ .}\eqno(5.5)$$
($T\sp {(n)}_{i_1\cdots i_n;j_1\cdots j_n}$ is defined in (A.1)).

Before coming to the consequences of this result, let us first discuss its
proof, which uses some of the identities given in the Appendix and proceeds by
induction. Obviously, (5.3) is true for $n=0$ by definition of
$T\sp {(0)}$. The explicit calculation of $W_1$ gives
$$\eqalign{W_1=&\ {\cal E}{\cal L}_1={\cal E}\bigl[F_1(\phi_i)\ W_0\bigr]\cr
=&\ M\sp {(1)}_{ij}\phi_{ij}\cr
=&\ {\rm tr}\ A\sp {(1)}=Tr\sp {(1)}(A\sp {(1)})\ ,\cr}\eqno(5.6)$$
in agreement with (5.3) (the last step uses (A.6)). Let us now assume that
(5.3) holds for the first $n\ts$ levels $(\ell=1,2,\cdots,n)$ of the
hierarchy. We then have to show that at the next level $(\ell=n+1)$ the
equation of motion $W_{n+1}$ computed from (5.2) does again agree with
(5.3). The latter calculation separates according to the following identity
$$\eqalign{{\cal E}\bigl[W(\phi_i,\phi_{ij})\ F(\phi_i)\bigr]=&
\bigl[{\cal E} W\bigr] F+2\bigl[{\pd W\over\pd\phi_i}
-{1\over 2}\pd_j\bigl({\pd W\over\pd\phi_{ij}}+{\pd W\over\pd\phi_{ji}}\bigr)
\bigr]\phi_{ik}{\pd F\over\pd\phi_k}+\cr
&+\bigl[W{\pd\sp 2 F\over\pd\phi_i\pd\phi_j}\phi_{ij}
-{\pd W\over\pd\phi_{ij}}{\pd\sp 2 F\over\pd\phi_k\pd\phi_l}
\phi_{ik}\phi_{jl}\bigr]\ ,}\eqno(5.7)$$
which is thus used here with $(W=W_n)$ given in (5.5) and $(F=F_{n+1})$.
Since $W_n$ is the equation of motion for the Lagrangian
$({\cal L}_n=F_n\ W_{n-1})$ which does not depend explicitly on the field
$\phi$, $W_n$ is simply a surface term. Consequently$\sp {[9,1]}$ its equation
of motion $({\cal E}W_n)$ -- the first term in (5.7) -- vanishes identically.
In fact, we have
$$\eqalign{{\cal E}W_n=&\ \pd_i\bigl[{\pd W_n\over\pd\phi_i}
-\pd_j{\pd W_n\over\pd\phi_{ij}}\bigr]\cr
=&\ \pd_i\bigl[{\pd W_n\over\pd\phi_i}
-{1\over 2}\pd_j\bigl({\pd W_n\over\pd\phi_{ij}}
+{\pd W_n\over\pd\phi_{ji}}\bigr)\bigr]\ ,}\eqno(5.8)$$
so that the equation of motion $({\cal E}W_n)$ is also the divergence of one
of the factors in the second term in (5.7). Actually, this factor itself
vanishes identically, as we now discuss.

First, consider the terms
$$\eqalign{\pd_j&\bigl[{\pd W_n\over\pd\phi_{ij}}
+{\pd W_n\over\pd\phi_{ji}}\bigr]=\cr
&=\bigl[{\pd\sp 2 W_n\over\pd\phi_{ij}\phi_k}
+{\pd\sp 2 W_n\over\pd\phi_{ji}\phi_k}\bigr]\phi_{jk}
+\bigl[{\pd\sp 2 W_n\over\pd\phi_{ij}\pd\phi_{kl}}
+{\pd\sp 2 W_n\over\pd\phi_{ji}\pd\phi_{kl}}\bigr]\phi_{jkl}
\ .\cr}\eqno(5.9)$$
Using the fact that $W_n$ is symmetric in its arguments
$M\sp {(p)}_{ij}\ (p=1,2,\cdots,n\ts)$ and that the symbol
$\epsilon_{i_1\cdots i_d}$ is fully antisymmetric, an explicit calculation
shows that the last two terms in (5.9) vanish:
$${\pd\sp 2 W_n\over\pd\phi_{ij}\pd\phi_{kl}}\phi_{jkl}=0\ ,\qquad
{\pd\sp 2 W_n\over\pd\phi_{ji}\pd\phi_{kl}}\phi_{jkl}=0\ .\eqno(5.10)$$
Similarly, using the same properties and the fact that
${\pd M\sp {(p)}_{ij}}/{\pd\phi_k}$ is actually symmetric in all three indices
$i,j,k$, an explicit analysis shows that the first two terms in (5.9) lead to
$${\pd\sp 2 W_n\over\pd\phi_{ij}\pd\phi_{k}}\phi_{jk}=
{\pd W_n\over\pd\phi_i}\ ,
\qquad{\pd\sp 2 W_n\over\pd\phi_{ji}\pd\phi_{k}}\phi_{jk}=
{\pd W_n\over\pd\phi_i}
\ .\eqno(5.11)$$
{}From the two results (5.10) and (5.11), it thus follows that we have indeed
$${\pd W_n\over\pd\phi_i}-{1\over 2}\pd_j\bigl[{\pd W_n\over\pd\phi_{ij}}
+{\pd W_n\over\pd\phi_{ji}}\bigr]=0\ .\eqno(5.12)$$

Therefore, using (5.7), (5.8) and (5.12), the calculation of $W_{n+1}$
reduces to
$$W_{n+1}=W_n M\sp {(n+1)}_{ij}\phi_{ij}
-{\pd W_n\over\pd\phi_{ij}}M\sp {(n+1)}_{kl}\phi_{ik}\phi_{jl}
\ .\eqno(5.13)$$
However, it should be clear that the net effect of the last term in this
relation is to replace each occurence of
$(A\sp {(p)}_{ij}=M\sp {(p)}_{ik}\phi_{kj})\ (p=1,2,\cdots,n\ts)$
in $W_n$ with
$(A\sp {(p)}_{ik}A\sp {(n+1)}_{kj})$, namely
$$\eqalign{W_{n+1}=&\ T\sp {(n)}_{i_1\cdots i_n;j_1\cdots j_n}
A\sp {(1)}_{i_1 j_1}\cdots A\sp {(n)}_{i_n j_n}A\sp {(n+1)}_{k k}-\cr
&-\sum_{p=1}\sp n T\sp {(n)}_{i_1\cdots i_n;j_1\cdots j_n}
A\sp {(1)}_{i_1 j_1}\cdots A\sp {(p-1)}_{i_{p-1}j_{p-1}}
A\sp {(p)}_{i_p k}A\sp {(n+1)}_{k j_p}A\sp {(p+1)}_{i_{p+1}j_{p+1}}
\cdots A\sp {(n)}_{i_n j_n}\cr
=&\ \bigl[T\sp {(n)}_{i_1\cdots i_n;j_1\cdots j_n}\delta_{i_{n+1}j_{n+1}}\cr
&-\sum_{p=1}\sp n T\sp {(n)}_{i_1\cdots i_n;j_1\cdots j_{p-1}j_{n+1}
j_{p+1}\cdots j_n}\delta_{i_{n+1}j_p}\bigr]
A\sp {(1)}_{i_1 j_1}\cdots A\sp {(n+1)}_{i_{n+1}j_{n+1}}\ .}\eqno(5.14)$$
Using now the identity (A.3), one finally obtains
$$\eqalign{W_{n+1}=&\ T\sp {(n+1)}_{i_1\cdots i_{n+1};j_1\cdots j_{n+1}}
A\sp {(1)}_{i_1 j_1}\cdots A\sp {(n+1)}_{i_{n+1}j_{n+1}}\cr
=&\ {(n+1)!}\ Tr\sp {(n+1)}(A\sp {(1)},\cdots,A\sp {(n+1)})\ ,\cr}
\eqno(5.15)$$
in agreement with (5.3). The iterative proof is thereby completed,
and the main theorem of this paper established.

Given the result (5.3), let us consider its consequences. It is clear
that $W_n$ is symmetric in the arguments $A\sp {(p)}$ or
$M\sp {(p)}\ (p=1,2,\cdots,n\ts)$.  Namely, the order in which the
multiplicative
factors $F_n(\phi_i)$ are introduced in the calculation of $W_n$ is
irrelevant;
$W_n$ is fully symmetric in these functions. Moreover, according to the
multiplication theorem (A.16), the dependence of $W_n$ on these functions
and on the second derivatives $\phi_{ij}$ separates as
$$W_n={n!}\ L\sp {(n)}_{i_{n+1}\cdots i_d;j_{n+1}\cdots j_d}
(M\sp {(1)},\cdots,M\sp {(n)})\ L\sp {(n)}_{j_{n+1}\cdots j_d;
i_{n+1}\cdots i_d}(\phi_{ij})\ .\eqno(5.16)$$
In particular, at level $(\ell=d\ts)$ we obtain
$$W_d={d!}\ Tr\sp {(d)}(M\sp {(1)},\cdots,M\sp {(d)})\ {\det \phi_{ij}}\ ,
\eqno(5.17)$$
showing that the dependence on the functions
$F_n(\phi_i)\ (n=1,2,\cdots,d\ts)$
factorizes at this level, leading to the following {\it universal} equation
of motion for the Lagrangian
${\cal L}_d$ independently of the factors $F_n(\phi_i)\ (n=1,2,\cdots,d\ts)$
$${\det \phi_{ij}}=0\ .\eqno(5.18)$$
Since the choice for the functions $F_n(\phi_i)\ (n=1,2,\cdots,d\ts)$ is
totally arbitrary, (5.18) is indeed an example of an equation of motion
admitting an infinite number of inequivalent Lagrangians. Moreover, since
$W_d$ is always a surface term -- being the equation of motion for a
Lagrangian without an explicit dependence on the field $\phi$ --, one also
concludes that there is an infinite number of inequivalent conserved
currents for the universal equation (5.18), suggesting its possible
integrability.

The result (5.17) also implies that the hierarchy terminates at that level.
Indeed, given any Lagrangian of the form
$${\cal L}(\phi_i,\phi_{ij})=F(\phi_i)\ {\det \phi_{ij}}\ ,\eqno(5.19)$$
as is ${\cal L}_{d+1}$, it is easily shown using (5.7) that its equation of
motion $({\cal E}{\cal L})$ vanishes identically. Hence, the universal
equation
(5.18) is also the last non trivial equation of motion for the hierarchy;
the recursive construction in (5.1) and (5.2) always leads to
$$W_n=0\ ,\qquad{\rm{for}}\ n\geq{d+1}\ ,\eqno(5.20)$$
which of course agrees implicitly with (5.3) since the functions
$Tr\sp {(n)}$ are defined only for $(n\leq d\ts)$.

In conclusion, we have established that the generic hierarchy indeed
satisfies all the properties i), iib), iii) - v) described in the
introduction, producing
the universal equation (5.18) or (1.1) at level $(\ell=d\ts)$. The generic
hierarchy is thus an example of a {\it finite} Euler hierarchy with a
{\it universal} equation of motion at the last level. In fact, in order for
the hierarchy to terminate after a finite number of iterations and to lead
to a universal equation of motion at the last level, it is crucial that the
factors $F_n(\phi_i)$ be functions only of the first derivatives of
the field  $\phi$. Were these functions to also depend on the field itself,
or on its second derivatives, the above properties would not follow. Namely,
at each successive level of the iterative procedure, derivatives of the field
of higher and higher order would be produced, never leading to a finite Euler
hierarchy nor to a universal equation of motion. It is quite remarkable
indeed that simply by taking for the multiplicative factors $F_n$ totally
arbitrary functions of first derivatives only, the Euler hierarchy only
involves first and second derivatives of the field, and terminates after a
finite number of steps with a universal equation of motion which is moreover
most probably an integrable system.

The above results generalise those of Ref.[1] corresponding to a situation
where all functions $F_n(\phi_i)$ are taken to be identical. Specifically,
consider the iterative procedure defining the generic hierarchy, with the
choice
$$F_n(\phi_i)=F_0(\phi_i)\ ,\quad n=1,2,\ldots\ ,\eqno(5.21)$$
where $F_0(\phi_i)$ is totally arbitrary. Then from (5.5) and (5.20),
it easily follows that
$$\sum_{n=0}\sp {\infty}{1\over n!}\lambda\sp n W_n=
{\det \bigl[\delta_{ij}+\lambda A\sp {(0)}_{ij}\bigr]}\ ,\eqno(5.22)$$
with of course
$$A\sp {(0)}_{ij}=M\sp {(0)}_{ik}\phi_{kj}\ ,\qquad
M\sp {(0)}_{ij}={\pd\sp 2 F_0\over\pd\phi_i\pd\phi_j}\ .\eqno(5.23)$$
The result (5.22) is essentially one of the two main theorems of Ref.[1],
which the present paper thus generalises.

To conclude this section, let us discuss the generalisation of the second
main theorem of Ref.[1] leading to the Bateman hierarchy. Namely, consider
again the above iterative procedure defining the generic hierarchy but now
with factors
$F_n(\phi_i)$ which are homogeneous functions of weight $(\alpha=1)$ but are
otherwise arbitrary. This situation corresponds to the Bateman hierarchy.
Thus, we now also have
$$F_n(\lambda\phi_i)=\lambda F_n(\phi_i)\ ,\eqno(5.24)$$
leading to the further properties
$$\eqalignno{M\sp {(n)}_{ij}(\lambda\phi_i)&=
\ \lambda\sp {-1}M\sp {(n)}_{ij}(\phi_i)\ ,&(5.25a)\cr
M\sp {(n)}_{ij}(\phi_i)\phi_j=0\quad&\Rightarrow
\quad{\det M\sp {(n)}_{ij}}=0\ .&(5.25b)\cr}$$
As we know from sect.2, the equations of motion $({\cal E}F_n)$ are then
invariant under arbitrary field redefinitions of $\phi$. Actually, this
general covariance property then extends to the whole hierarchy. Indeed,
consider an arbitrary transformation
$$\varphi=F(\phi)\ ,\eqno(5.26a)$$
so that
$$\varphi_i=F\sp {\prime}(\phi)\phi_i\ ,\quad
\varphi_{ij}=F\sp {\prime}(\phi)\phi_{ij}
+F\sp {\prime\prime}(\phi)\phi_i\phi_j\ .\eqno(5.26b)$$
It then follows that
$$\eqalign{A\sp {(n)}_{ij}(\varphi_i,\varphi_{ij})=&
\ M\sp {(n)}_{ik}(\varphi_i)\varphi_{kj}\cr
=&\ {F\sp {\prime}(\phi)}\sp {-1}M\sp {(n)}_{ik}(\phi_i)
\bigl[F\sp {\prime}(\phi)\phi_{kj}+F\sp {\prime\prime}
(\phi)\phi_k\phi_j\bigr]\cr
=&\ A\sp {(n)}_{ij}(\phi_i,\phi_{ij})\ ,\cr}\eqno(5.27)$$
where the second line uses (5.25b). Consequently, we have that
$$W_n(\varphi_i,\varphi_{ij})=W_n(\phi_i,\phi_{ij})\ ,\eqno(5.28)$$
showing that all equations of motion of the hierarchy are then also invariant
under arbitrary field redefinitions. In other words, all Lagrangians
${\cal L}_n$ constructed from the hierarchy define classical topological
field theories of one field in $d\ts$ dimensions, provided the multiplicative
factors $F_n(\phi_i)$ themselves define such theories, namely are weight one
homogeneous functions of $\phi_i$. In fact, all Lagrangians ${\cal L}_n$
constructed from the generic hierarchy then obey the homogeneity property
(2.9) with weight $(\alpha=1)$. The Bateman hierarchy is thus a hierarchy
of classical topological field theories.

More specifically, the homogeneous functions $F_n(\phi_i)$ may always be
represented as
$$F_n(\phi_i)={\phi_d}\ G\sp {(n)}({\phi_\alpha\over\phi_d})\ ,\eqno(5.29)$$
by an appropriate choice of labelling of the coordinates $x_i$. Here, the
quantities $G\sp {(n)}$ are arbitrary functions of their $(d-1)$
arguments and $(\alpha,\beta)$ stand for the indices labelling the
first $(d-1)$ coordinates
among the coordinates $x_i$. Given (5.29), it then follows that
$$M\sp {(n)}_{ij}(\phi_i)=\pmatrix{{G\sp {(n)}_{\alpha\beta}\over\phi_d}&
{G\sp {(n)}_{\alpha\beta}\over\phi_d}(-{\phi_\beta\over\phi_d})\cr
(-{\phi_\alpha\over\phi_d}){G\sp {(n)}_{\alpha\beta}\over\phi_d}&
(-{\phi_\alpha\over\phi_d}){G\sp {(n)}_{\alpha\beta}\over\phi_d}
(-{\phi_\beta\over\phi_d})\cr}\ .\eqno(5.30)$$
Here, $G\sp {(n)}_{\alpha\beta}$ is the matrix of second derivatives of the
function $G\sp {(n)}$ with respect to its arguments -- these arguments
taking the
values $({\phi_\alpha}/{\phi_d})$ in (5.30) --, and $(i,\alpha)$ and
$(j,\beta)$ are line and column indices respectively. Using the degeneracy
theorem (A.19) in the Appendix, one then concludes that the generalised
determinant function $Tr\sp {(d)}(M\sp {(1)},\cdots,M\sp {(d)})$ vanishes
identically, implying that
$$W_d(\phi_i,\phi_{ij})=0\ .\eqno(5.31)$$
In other words, the Bateman hierarchy terminates one step earlier than the
generic hierarchy, namely at level $(\ell=d-1)$. The last non trivial
equation
of motion of the Bateman hierarchy is thus $W_{d-1}$. From (5.16) and the
results (A.13) and (A.25), one finds explicitly
$$W_{d-1}=-{(d-1)!\over{\phi_d}\sp {d+1}}\ Tr\sp {(d-1)}_{(d-1)}
(G\sp {(1)}_{\alpha\beta},\cdots,G\sp {(d-1)}_{\alpha\beta})
\ {\det\pmatrix{0&\phi_j\cr\phi_i&\phi_{ij}\cr}}\ .\eqno(5.32)$$
Hence here again, the dependence on the choice of functions $F_n$ for the last
non trivial equation of motion of the Bateman hierarchy factorizes out,
leading to the universal Bateman equation of motion at level $(\ell=d-1)$
$${\det\pmatrix{0&\phi_j\cr\phi_i&\phi_{ij}\cr}}=0\ .\eqno(5.33)$$

These are thus the results promised in the introduction in the case of the
Bateman hierarchy. They also generalise the conclusions of Ref.[1] which were
obtained in the case when all homogeneous weight one functions $F_n$
are equal. Note that any field configuration obeying the relation
$$\phi_{ij}=S_i\phi_j+S_j\phi_i\eqno(5.34)$$
for some coefficients $S_i$, is always a solution to {\it all} equations of
motion of the Bateman hierarchy -- generalising a similar statement
in Ref.[1].
Indeed, when (5.34) is used in the expression for $W_n$, each of the factors
$\phi_i$ is always contracted with one of the matrices
$M\sp {(p)}_{ij}\ (p=1,2,\cdots,n\ts)$, leading to a vanishing contribution
since $\phi_i$ is a zero mode of these matrices (see (5.25b)).
In particular, configurations $\phi(x_i)$ implicitly defined by the equation
$${x_i}{F_i(\phi)}=c\ ,\eqno(5.35)$$
with $F_i(\phi)$ being arbitrary functions and $c$ an arbitrary
constant, always obey the identity (5.34). Thus, (5.35) defines$\sp {[1]}$
implicit solutions to {\it all} the equations of motion of the Bateman
hierarchy, including the Bateman equation (5.33).

Finally, one may wonder whether by imposing further conditions on
the functions $F_n(\phi_i)$, the Bateman hierarchy itself would not
terminate a step earlier in the same way that the generic hierarchy
terminates at level
$(\ell=d-1)$ when the functions $F_n(\phi_i)$ are homogeneous and of weight
one. Actually, the explicit result (5.32) shows that $W_{d-1}$ itself
vanishes whenever the generalised determinant which is the prefactor in that
expression vanishes. From the discussion around (A.19) in the Appendix, it
follows that such a situation indeed occurs provided the functions
$G\sp {(n)}$ introduced in (5.29) are themselves homogeneous functions of
weight one, namely
$$G\sp {(n)}(t_\alpha)=t_{d-1}\ H\sp {(n)}({t_1\over t_{d-1}},
\cdots,{t_{d-2}\over t_{d-1}})\ .\eqno(5.36)$$
However, this choice implies
$$F_n(\phi_i)=\phi_{d-1}\ H\sp {(n)}({\phi_1\over\phi_{d-1}},
\cdots,{\phi_{d-2}\over\phi_{d-1}})\ ,\eqno(5.37)$$
thus showing that one is then actually dealing with a Bateman hierarchy
for a single field in $(d-1)$ dimensions. The last coordinate $x_d$ plays no
r$\hat{\rm o}$le
in the dynamical evolution of the field theory, since the functions
$F_n$, hence also the successive Lagrangians in the hierarchy, are
independent of derivatives of the field with respect to that coordinate.
Therefore, weight one homogeneity is the only possible restriction on the
functions $F_n$ leading to a truncation of the generic hierarchy at
an earlier stage. Correspondingly, one obtains the Bateman hierarchy with
the universal Bateman equation (5.33) at the last non trivial level.

\vskip 20pt
\leftline{\bf 6. Finite Euler Hierarchies for Many Fields}
\vskip 20pt
Given the generic finite Euler hierarchy for one field with a universal
equation of motion at the last level, it is now possible to construct other
such hierarchies for one or more fields. One obvious way is to apply the
C- or R-maps to the generic hierarchy for one field depending on $d\ts$
coordinates. Using the C-map, one obtains a finite Euler hierarchy of
classical topological field theories for one field in $(d+1)$ dimensions.
Actually, this hierarchy is
precisely the Bateman hierarchy described at the end of the previous section,
with the Bateman equation (5.33) as its universal equation at level
$(\ell=d\ts)$. Using the R-map, one obtains$\sp {[2]}$ a finite Euler
hierarchy
of reparametrisation invariant field theories for $(d+1)$ fields in $d\ts$
dimensions, whose universal equation at level $(\ell=d\ts)$ was given in
(1.3) (there are of course $(d+1)$ equations of motion at each level of this
hierarchy, but the Ward identities of reparametrisation invariance imply that
there is actually only one independent equation at each level. At the last
level, this is the universal equation (1.3)).

However, it is possible to consider a somewhat wider framework which includes
the hierarchies just mentioned as particular cases, and allows to have more
than one field still in the context of the generic hierarchy. The idea is
simply to consider the generic hierarchy with the original field $\phi(x_i)$
now being actually a linear combination of a collection of fields
$\phi\sp a(x_i)$, namely $(\phi(x_i)=\phi\sp a(x_i)\lambda\sp a)$.
The coefficients $\lambda\sp a$ are totally arbitrary, and it is assumed
that whatever the equations
of motion, they are expanded in terms of these coefficients and have to hold
independently of their values. The situation discussed above is recovered
simply by setting the then single coefficient $(\lambda=1)$, namely by
absorbing it in the normalisation of the field $\phi$ itself. For this reason,
the hierarchies described in the previous paragraph are not
discussed separately here.

First, let us thus discuss the generic case of $D\ts$ fields
$\phi\sp a(x_i)$
functions of $d\ts$ coordinates $x_i$. Consider a collection of arbitrary
functions $F_n(t_i)\ (n=1,2,\ldots)$ of $d\ts$ arguments denoted $t_i$
for our present purpose, with the symmetric matrix functions
$$M\sp {(n)}_{ij}(t_i)={\pd\sp 2 F_n\over\pd t_i\pd t_j}(t_i)\ .\eqno(6.1)$$
Define now the iterative construction of the following Euler hierarchy. Its
Lagrangians for $(n=1,2,\ldots)$ are given by
$${\cal L}_{(n;a_1\cdots a_{n-1})}(\phi\sp a_i,\phi\sp a_{ij};\lambda\sp a)=
F_n(\phi\sp a_i\lambda\sp a)W_{(n-1;a_1\cdots a_{n-1})}
(\phi\sp a_i,\phi\sp a_{ij};\lambda\sp a)\ ,\eqno(6.2a)$$
where
$$W_0=1\ ,\eqno(6.2b)$$
and
$$W_{(n;a_1\cdots a_n)}(\phi\sp a_i,\phi\sp a_{ij};\lambda\sp a)=
{\cal E}_{\phi\sp {a_n}}{\cal L}_{(n;a_1\cdots a_{n-1})}
\bigl[\phi\sp a\bigr]\ ,\quad{\rm{for}}\ n=1,2,\ldots\ .\eqno(6.2c)$$
By relating this construction to that of the generic hierarchy for the field
$(\Phi=\phi\sp a\lambda\sp a)$ with the functions $F_n(\Phi_i)$, it follows
that we actually have the explicit result
$$W_{(n;a_1\cdots a_n)}(\phi\sp a_i,\phi\sp a_{ij};\lambda\sp a)=
{n!}\ \lambda\sp {a_1}\cdots\lambda\sp {a_n}Tr\sp {(n)}
(A\sp {(1)},\cdots,A\sp {(n)})\ ,\eqno(6.3)$$
for $(n=0,1,\ldots)$, with
$$A\sp {(p)}_{ij}=M\sp {(p)}_{ik}
(\phi\sp a_i\lambda\sp a)\phi\sp a_{kj}\lambda\sp a\ ,\qquad
p=1,2,\cdots,n\ .\eqno(6.4)$$
In particular, this implies that the hierarchy defined in (6.2) terminates at
level $(\ell=d\ts)$, with the universal equations
$${\det(\phi\sp a_{ij}\lambda\sp a)}=0\ .\eqno(6.5)$$
As before, the dependence of the last non trivial equations of motion
$W_{(d;a_1\cdots a_d)}$ on the functions $F_n$ factors out, leaving (6.5) as
the universal equations for this hierarchy. By expansion in the coefficients
$\lambda\sp a$, one thus obtains $d+D-1\choose D-1$ equations of motion for
the $D\ts$ fields $\phi\sp a$. Even though this may seem to be an
overdetermined
system -- except when $(D=1)$ or $(d=1)$ --, it is easy to see that it
nevertheless possesses non trivial solutions (some are discussed in the next
section). Moreover, note that there is essentially only one independent
Lagrangian at every level of the hierarchy, since the free indices $a_k$
attached to the Lagrangians (6.2a) are simply carried by the coefficients
$\lambda\sp {a_k}$ as overall factors.

Obviously, the generic hierarchy is indeed recovered from the expressions
above in the case $(D=1)$ by setting $(\lambda=1)$. The other situation
were the number of universal equations of motion is equal to the number
of fields, namely $(d=1)$, corresponds to a hierarchy terminating after
one step with the equations
$$\phi\sp a_{xx}=0\ ,\qquad a=1,2,\cdots,D\ .\eqno(6.6)$$
The general solution to these equations is obvious.

It is straightforward to apply the C-map to the generic hierarchy for many
fields defined above. As we know, this leads to a hierarchy of classical
topological field theories, which is constructed as follows (the notation is
that of sect.3). Consider $(D=p)$ fields $\phi\sp a(x_i)$ dependent on
$(d=p+q)$ coordinates $(x_i=(x_\alpha,y\sp a))$ (there are thus $q\ts$
coordinates $x_\alpha$ and $p\ts$ coordinates $y\sp a$), and introduce
the matrix
$$M_a{}\sp b={\pd\phi\sp b\over\pd y\sp a}\ .\eqno(6.7)$$
Consider now functions $F_n\ (n=1,2,\ldots)$ given by
$$F_n(\phi\sp a_i;\lambda\sp a)=({\det M_a{}\sp b})
\ G_n(-{\pd\phi\sp a\over\pd x_\alpha}(M\sp {-1})_a{}\sp b\lambda\sp b)
\ ,\eqno(6.8)$$
where $G_n(t_\alpha)$ are arbitrary functions of $q\ts$ arguments $t_\alpha$,
with the symmetric matrix functions
$$M\sp {(n)}_{\alpha\beta}(t_\alpha)=
{\pd\sp 2 G_n\over\pd t_\alpha\pd t_\beta}(t_\alpha)\ .\eqno(6.9)$$
Define then iteratively the following Euler hierarchy. Its Lagrangians for
$(n=1,2,\ldots)$ are given by
$$\eqalign{{\cal L}_{(n;a_1\cdots a_{n-1})}
&(\phi\sp a_i,\phi\sp a_{ij};\lambda\sp a)=\cr
&=F_n(\phi\sp a_i;\lambda\sp a)({\det M_a{}\sp b})\sp {-1}M_{a_{n-1}}{}\sp b
\ W_{(n-1;a_1\cdots a_{n-2}b)}
(\phi\sp a_i,\phi\sp a_{ij};\lambda\sp a)\ ,\cr}
\eqno(6.10a)$$
where
$$W_0={\det M_a{}\sp b}\ ,\eqno(6.10b)$$
and
$$W_{(n;a_1\cdots a_n)}(\phi\sp a_i,\phi\sp a_{ij};\lambda\sp a)=
{\cal E}_{\phi\sp {a_n}}{\cal L}_{(n;a_1\cdots a_{n-1})}
\bigl[\phi\sp a\bigr]\ ,\quad{\rm{for}}\ n=0,1,\ldots\ .\eqno(6.10c)$$
{}From the C-map applied to the hierarchy defined in (6.2), it follows that
we have in fact for $(n=0,1,\ldots)$
$$\eqalign{W_{(n;a_1\cdots a_n)}&
(\phi\sp a_i,\phi\sp a_{ij};\lambda\sp a)=\cr
&=(-1)\sp n{n!}\ ({\det M_a{}\sp b})
\ \lambda\sp {a_1}\cdots\lambda\sp {a_{n-1}}
(M\sp {-1})_{a_n}{}\sp b\lambda\sp b\ Tr\sp {(n)}
(A\sp {(1)},\cdots,A\sp {(n)})\ ,\cr}
\eqno(6.11)$$
with
$$A\sp {(p)}_{\alpha\beta}=M\sp {(p)}_{\alpha\gamma}
(Y\sp a_\alpha\lambda\sp a)
Y\sp a_{\gamma\beta}\lambda\sp a\ ,\eqno(6.12)$$
where the quantities $Y\sp a_\alpha$ and
$Y\sp a_{\alpha\beta}$ are given in (3.8)
and (3.9). As already mentioned, the construction in (6.10) leads to a
hierarchy of classical topological field theories. In particular, note how
the functions $F_n$ in (6.8) obey the homogeneity property (2.9) with weight
$(\alpha=1)$, ensuring that their equations of motion are generally covariant
under arbitrary field redefinitions. The homogeneity property (2.9) with
weight $(\alpha=1)$ of course extends to the whole hierarchy itself.
Also, note that there is in fact essentially only one independent
Lagrangian at each level of the hierarchy.

{}From (6.11), it follows that the hierarchy (6.10) is again a finite one,
terminating at the level $(\ell=q=d-D)$ with equations of motion which, up
to a factor depending on the functions $F_n$, are universal and given by
$${\det(Y\sp a_{\alpha\beta}\lambda\sp a)}=0\ .\eqno(6.13)$$
Using the identity
$${\det(X_{\alpha\beta}-A_{\alpha a}B_{a\beta}
-B_{a\alpha}A_{\beta a}+A_{\alpha a}Y_{ab}A_{\beta b})}=
{\det\pmatrix{0_{ab}&A_{\beta a}&\delta_{ab}\cr
A_{\alpha b}&X_{\alpha\beta}&B_{b\alpha}\cr
\delta_{ab}&B_{a\beta}&Y_{ab}\cr}}\ ,\eqno(6.14)$$
(here, $(a,\alpha)$ and $(b,\beta)$ are line and column indices
respectively), the equations implied by (6.13) are equivalent to
$${\det\pmatrix{0&\phi\sp b_i\cr\phi\sp a_j&\phi\sp c_{ij}\lambda\sp c\cr}}
=0\ ,\eqno(6.15)$$
(here, $(i,a)$ and $(j,b)$ are line and column indices respectively). The
hierarchy (6.10) thus leads to the generally covariant equations of
motion (1.5) as universal equations of motion arising at level $(\ell=d-D)$.
This establishes the conjecture of Ref.[1]. Clearly, the hierarchy (6.10)
is a generalisation of the Bateman hierarchy for one field. The latter is
indeed recovered when $(D=1=p)$ and setting $(\lambda=1)$ in the above
expressions. Note how the functions $F_n$ in (6.8) are then precisely of the
form (5.29) used in the construction of the Bateman hierarchy as a
particular reduction of the generic hierarchy.

As pointed out in the introduction, the set of $d-1\choose D-1$ equations is
in general overdetermined, except when $(D=1)$ or $(D=d-1)$. These two
special cases of course correspond through the C-map to the two
distinguished cases $(D=1)$ and $(d=1)$ mentioned above in the context of the
generic hierarchy for many fields. Here, the situation for $(D=1)$ is that of
the Bateman hierarchy. On the other hand, the situation with $(D=d-1)$
corresponds precisely to universal equations already given in Ref.[1].
Namely, when $(D=d-1)$, the hierarchy (6.10) terminates after only one
iteration with the universal equations of motion
$$J_i J_j\phi\sp a_{ij}=0\ ,\eqno(6.16)$$
where the Jacobians $J_i$ are defined by
$${\det\pmatrix{\phi\sp b_i&\zeta_i\cr}}=\zeta_i J_i\ ,\eqno(6.17)$$
(here, $i$ and $b$ are line and column indices respectively). The solutions
to these equations are obtained from those of (6.6) through the C-map.

Finally, let us describe the result of the R-map applied to the generic
hierarchy for more than one field defined in (6.2). As we know,
one then obtains hierarchies of reparametrisation invariant field theories.
Using the notations of sect.4, consider $(D=p+q)$ fields
$(\phi\sp a(x_j)=(\phi\sp \mu(x_j),z_i(x_j)))$ dependent on $(d=q)$
coordinates $x_i$ (there are thus $p\ts$ fields $\phi\sp \mu$ and $q\ts$
fields $z_i$), and define the matrix
$$M_i{}\sp j={\pd z_j\over\pd x_i}\ .\eqno(6.18)$$
Consider now a collection of functions $F_n\ (n=1,2,\ldots)$ given by
$$F_n(\phi\sp a_i;\lambda\sp \mu)=({\det M_i{}\sp j})
\ G_n((M\sp {-1})_i{}\sp j
{\pd\phi\sp \mu\over\pd x_j}\lambda\sp \mu)\ ,\eqno(6.19)$$
where $G_n(t_i)$ are arbitrary functions of $(d=q)$ arguments $t_i$,
with the symmetric matrix functions
$$M\sp {(n)}_{ij}(t_i)={\pd\sp 2 G_n\over\pd t_i\pd t_j}(t_i)
\ .\eqno(6.20)$$
Define then iteratively the following hierarchy. The Lagrangians for
$(n=1,2,\ldots)$ are given by
$${\cal L}_{(n;\mu_1\cdots\mu_{n-1})}
(\phi\sp a_i,\phi\sp a_{ij};\lambda\sp \mu)=
F_n(\phi\sp a_i;\lambda\sp \mu)({\det M_i{}\sp j})\sp {-1}
W_{(n-1;\mu_1\cdots\mu_{n-1})}
(\phi\sp a_i,\phi\sp a_{ij};\lambda\sp \mu)\ ,\eqno(6.21a)$$
where
$$W_0={\det M_i{}\sp j}\ ,\eqno(6.21b)$$
and
$$W_{(n;\mu_1\cdots\mu_n)}(\phi\sp a_i,\phi\sp a_{ij};\lambda\sp \mu)=
{\cal E}_{\phi\sp {\mu_n}}{\cal L}_{(n;\mu_1\cdots\mu_{n-1})}
\bigl[\phi\sp a\bigr]\ ,\quad{\rm{for}}\ n=0,1,\ldots\ .\eqno(6.21c)$$
{}From the R-map, one then finds the explicit representation for
$(n=0,1,2,\ldots)$
$$W_{(n;\mu_1\cdots\mu_n)}(\phi\sp a_i,\phi\sp a_{ij};\lambda\sp \mu)=
{n!}\ ({\det M_i{}\sp j})\ \lambda\sp {\mu_1}\cdots\lambda\sp {\mu_n}
\ Tr\sp {(n)}(A\sp {(1)},\cdots,A\sp {(n)})\ ,\eqno(6.22)$$
where
$$A\sp {(p)}_{ij}=M\sp {(p)}_{ik}(Y\sp \mu_i\lambda\sp \mu)
Y\sp \mu_{kj}\lambda\sp \mu\ ,\eqno(6.23)$$
with $Y\sp \mu_i$ and $Y\sp \mu_{ij}$ given in (4.6) and (4.7). As already
mentioned,
the construction (6.21) leads to a Euler hierarchy of reparametrisation
invariant field theories. In particular, the functions $F_n$ obey the
homogeneity property (2.3) with weight $(\alpha=1)$, ensuring
reparametrisation invariance of the field theories they define. The property
(2.3) with weight $(\alpha=1)$ of course extends to the whole hierarchy. Note
that the hierarchy is defined only in terms of equations of motion for the
fields $\phi\sp \mu$. This is because the Ward identities of reparametrisation
invariance relate the equations of motion for the fields $z_i$ to those for
the fields $\phi\sp \mu$, the latter thus being the only independent ones.
Also, note that once again there is essentially only one independent
Lagrangian at each level of the hierarchy.

As before, (6.22) implies that the hierarchy (6.21) is also finite, and
terminates at level $(\ell=d=q)$ with equations of motion which, up to a
factor depending on the choice of functions $F_n$, are universal equations
with reparametrisation invariance. These universal equations are obtained
from
$${\det(Y\sp \mu_{ij}\lambda\sp \mu)}=0\ ,\eqno(6.24)$$
expanded in terms of the coefficients $\lambda\sp \mu$. For convenience,
let us introduce Jacobians $J$ and $J\sp \mu_i$ generated by the function
$${\det\pmatrix{\zeta&\zeta\sp j\cr
\phi\sp \mu_i\lambda\sp \mu&M_i{}\sp j\cr}}=
\zeta J+\zeta\sp i J\sp \mu_i\lambda\sp \mu\ ,\eqno(6.25)$$
where as usual $i$ and $j$ denote line and column indices respectively. The
explicit relations are
$$J={\det M_i{}\sp j}\ ,\qquad J\sp \mu_i=-({\det M_i{}\sp j})
(M\sp {-1})_i{}\sp j {\pd\phi\sp \mu\over\pd x_j}\ .\eqno(6.26)$$
Note that $J\sp \mu_i$ is the determinant of the matrix $M_i{}\sp j$ in
which the $i\sp {{\rm th}}$ line is replaced by the line
$({\pd\phi\sp \mu}/{\pd x_j})$. In
terms of these definitions, the universal equations (6.24) reduce to
$${\det\bigl[(J{\pd\sp 2\phi\sp \mu\over\pd x_i\pd x_j}+
J\sp \mu_k{\pd\sp 2 z_k\over\pd x_i\pd x_j})\lambda\sp \mu\bigr]}=0
\ .\eqno(6.27)$$
One thus obtains $({p+q-1\choose q}={D-1\choose d})$ independent universal
equations of motion for a system of $D\ts$ fields which is reparametrisation
invariant in $d\ts$ coordinates. Except when $(D=d+1)$ or $(d=1)$
-- corresponding through the R-map to the two cases when the universal
equations of motion of the generic hierarchy (6.2) are not overdetermined --,
the system of universal equations (6.27) is overdetermined since the
number of independent equations should be $(D-d\ts)$, when accounting for
reparametrisation invariance. Nevertheless, it is easy to see that there exist
non trivial solutions to these equations.

{}From the results above, the reparametrisation invariant hierarchy obtained
from the R-map acting on the generic hierarchy for one field is recovered when
$(D=d+1)$, i.e. $(p=1)$, and setting $(\lambda=1)$. The corresponding
hierarchy then terminates at level $(\ell=d\ts)$ with the universal
reparametrisation invariant equation of motion (1.3) first considered in
Ref.[2]. This is one of the two instances when the number of fields and
universal equations are equal, up to the Ward identities of reparametrisation
invariance. The other such instance is when $(d=1)$. In this case, the
hierarchy (6.21) terminates after one iteration, with the $(D-1)$ independent
universal equations of motion
$${\pd z\over\pd x}{\pd\sp 2\phi\sp \mu\over\pd x\sp 2}=
{\pd\phi\sp \mu\over\pd x}{\pd\sp 2 z\over\pd x\sp 2}\ .\eqno(6.28)$$
Consequently we then have$\sp {[2]}$
$${\pd\phi\sp a\over\pd x}{\pd\sp 2\phi\sp b\over\pd x\sp 2}=
{\pd\phi\sp b\over\pd x}{\pd\sp 2\phi\sp a\over\pd x\sp 2}\ ,
\quad{\rm{for\ all}}\ a,b\ ,
\eqno(6.29)$$
whose solutions are obtained from those of (6.6) through the R-map.

To conclude, let us emphasize the fact that the hierarchies constructed in
this section are generalisations of the finite Euler hierarchies with
universal equations discovered and conjectured in Refs.[1,2]. These
generalisations now apply to field theories for which the difference between
the numbers of fields and coordinates is totally arbitrary. The associated
universal equations are given in (6.5), (6.15) and (6.27) and are dual to one
another under the C- and R-maps. These equations include as particular
examples the cases (1.1), (1.2) and (1.3) as well as (6.6), (6.16) and (6.29),
corresponding to the finite Euler hierarchies considered
previously$\sp {[1,2]}$.
\vskip 20pt
\leftline{\bf 7. Examples of Solutions}
\vskip 20pt
As pointed out repeatedly, the universal equations of motion discussed above
possess an infinite number of conservation laws. Indeed, these equations
derive from an infinity of inequivalent Lagrangians, {\it i.e.} Lagrangians
not differing by surface terms, which do not explicitly depend on the fields
themselves but only on their derivatives. Hence, the corresponding equations
of motion take the form
$${\cal E}_a{\cal L}(\phi\sp a_i,\phi\sp a_{ij})=
\pd_i\bigl[{\pd{\cal L}\over\pd\phi\sp a_i}
-\pd_j{\pd{\cal L}\over\pd\phi\sp a_{ij}}\bigr]\ ,\eqno(7.1)$$
which are thus conservation laws for field configurations solving the
equations of motion. This property strongly suggests the possible
integrability of the universal equations, namely the existence of an
infinite number of conserved charges in involution for some symplectic
structure defining a Hamiltonian formulation for these systems.

This suggestion is also supported from a somewhat different point of view,
although not completely unrelated to the previous remark. All theories in
this paper are invariant under translations in the base space coordinates
$x_i$. Hence, by Noether's first theorem$\sp {[10]}$, there exists a
conserved current associated to this symmetry: the stress-tensor $T_{ij}$.
For Lagrangians ${\cal L}(\phi\sp a_i,\phi\sp a_{ij})$ depending only
on first and second derivatives of the fields, we have explicitly
$$T_{ij}=\phi\sp a_i{\pd{\cal L}\over\pd\phi\sp a_j}
+{1\over 2}\phi\sp a_{ik}
\bigl[{\pd{\cal L}\over\pd\phi\sp a_{jk}}
+{\pd{\cal L}\over\pd\phi\sp a_{kj}}\bigr]
-{1\over 2}\phi\sp a_i \pd_k
\bigl[{\pd{\cal L}\over\pd\phi\sp a_{jk}}
+{\pd{\cal L}\over\pd\phi\sp a_{kj}}\bigr]
-\delta_{ij}{\cal L}\ .\eqno(7.2)$$
(Note that for reparametrisation invariant theories obeying the homogeneity
property (2.3) with weight $(\alpha=1)$, this quantity vanishes identically).
The conservation equation for the stress-tensor is
$$\pd_j T_{ij}=\phi\sp a_i {\cal E}_a{\cal L}\ .\eqno(7.3)$$
Therefore, associated to the infinity of Lagrangians leading to the universal
equations of this paper, solutions to these equations possess the infinity of
conservation laws for the corresponding stress-tensors $T_{ij}$ whenever the
latter is non trivial.

It is not the purpose of this paper to address the issue of integrability,
neither for cases 1) -- 3) of the introduction nor for the more general
universal equations of the previous section. However, we present in this
section large classes of solutions, thereby providing circumstancial evidence
that we are indeed dealing with integrable systems. In the simplest cases, it
is possible to give the general solution to the equations -- sometimes taking
advantage of the C- and R-maps --, thus implicitly proving their
integrability. On the other hand, let us recall$\sp {[1]}$ how the
integrability of the two dimensional Bateman equation may be displayed
explicitly. Introducing$\sp {[5,1]}$ the variables
$$u(x_1,x_2)={\phi_1(x_1,x_2)\over\phi_2(x_1,x_2)}\ ,\eqno(7.4)$$
the two dimensional Bateman equation (1.2) reduces to the following non
linear equation for $u(x_1,x_2)$
$$u_1=\ u u_2\ .\eqno(7.5)$$
This is recognised as the KdV equation without the term in
$u_{222}$. Obviously, all integer powers of $u$ define
densities of conserved charges for this system, which are also in involution
for either of the two known symplectic structures for the KdV hierarchy.

Let us first concentrate on the simplest cases for which the general solution
may be constructed, in the case of the generic hierarchy for one field
and its C- and R-dual representations. Obviously, the Bateman equation in one
dimension is trivially solved: $\phi(x)$ is simply any constant. Next,
consider the universal equation (1.1) for the generic hierarchy in one
dimension, which is simply
$$\phi_{xx}=0\ .\eqno(7.6)$$
The general solution is thus
$$\phi(x)=\ {A}{x}+{B}\ ,\eqno(7.7)$$
where $A$ and $B$ are arbitrary constants. Using the R-map, we then directly
obtain the general solution to the reparametrisation invariant equation (1.3)
in the case of two fields $\phi\sp 1$ and $\phi\sp 2$ in one dimension.
The solution is
$$\phi\sp 1(x)=\ {A}{f(x)}+{B}\ ,\qquad \phi\sp 2(x)=f(x)\ ,\eqno(7.8)$$
where $f(x)$ is {\it any} function of the coordinate $x$, possibly constant,
and $A$ and $B$ are arbitrary constants. Actually, the condition (4.4)
-- necessary for the application of the R-map -- requires the function $f(x)$
to be non constant. The only solutions to (1.3) when $(d=1)$ which do not
obey (4.4) are precisely those for which $f(x)$ is constant. Hence, (7.8)
indeed defines the general solution to (1.3) in this case, with $f(x)$ being
any function.

Using the C-map, (7.7) also leads to the following implicit solutions to the
Bateman equation (1.2) in two dimensions
$$x_2=\ A(\phi){x_1}+B(\phi)\ ,\eqno(7.9)$$
where $A(\phi)$ and $B(\phi)$ are now arbitrary functions of $\phi$. The C-map
does not necessarily provide all solutions however, since the non degeneracy
condition (3.6) must be obeyed. In the present case, the solutions which are
not given by (7.9) are simply those which are independent of $x_2$.
Obviously, all solutions are obtained when (7.9) is extended to
$$x_1 F(\phi)+x_2 G(\phi)+H(\phi)=0\ ,\eqno(7.10)$$
where $F(\phi)$, $G(\phi)$ and $H(\phi)$ are arbitrary functions. Any choice
with $(G( \phi)\neq 0)$ leads back to (7.9), whereas taking $(G(\phi)=0)$
leads to {\it all} configurations $\phi(x_1)$ function of $x_1$ only, which
are indeed always solutions.

The results above thus completely describe the general solutions to all three
equations (1.1) -- (1.3) when $(d=1)$. These systems are clearly integrable,
since one of the members of this ``duality triangle'' is the two dimensional
Bateman equation, which is indeed an integrable system as was explained in
(7.5). Note also how (7.10) is equivalent to (1.6) in this case, showing that
(1.6) indeed defines the general solution to the Bateman equation in
two dimensions.

Let us now turn to the universal equation (1.1) of the generic hierarchy when
$(d=2)$. This is a particular reduction of the Plebanski equation$\sp {[3]}$
for self-dual gravity in four dimensions, which is known to be integrable.
Therefore, its dual pair -- the Bateman equation (1.2) in three dimensions
and the universal string equation (1.3) in three spacetime dimensions
-- are also integrable systems. In fact, all solutions to these equations
may be defined implicitly. Consider first the equation (1.1) in this case.
Introducing the new variable $(h(x_1,x_2)=\phi_2(x_1,x_2))$,
it follows$\sp {[3]}$
that if $\phi(x_1,x_2)$ solves (1.1) then $h(x_1,x_2)$ is a solution to the
{\it two dimensional Bateman equation}, whose general solutions are given
above. Considering then separately the two cases when either $h(x_1,x_2)$ is
independent of $x_2$ or is given as in (7.9), and using the fact that any
function of a solution to the Bateman equation is again a solution, one
concludes that all solutions to (1.1) when $(d=2)$ are obtained from the
following two relations
$$\phi(x_1,x_2)=x_1 F(h)+x_2 G(h)+H(h)\ ,\eqno(7.11)$$
and
$$x_1{{\rm d}F\over{\rm d}h}(h)+x_2{{\rm d}G\over{\rm d}h}(h)
+{{\rm d}H\over{\rm d}h}(h)=\ 0\ .\eqno(7.12)$$
Namely, (7.12) defines an implicit solution for $h(x_1,x_2)$ which is then
used in (7.11) to obtain the solution for $\phi(x_1,x_2)$. Here, $F(h)$,
$G(h)$ and $H(h)$ are totally arbitrary functions.

Using the R-map, one then also obtains the general solution to the universal
string equation (1.3) in a three dimensional spacetime. The three fields
$\phi\sp a(x_1,x_2)\ (a=1,2,3)$ functions of the coordinates $(x_1,x_2)$
are then given by
$$\eqalign{\phi\sp 1(x_1,x_2)=&f_1(x_1,x_2)F(h)+f_2(x_1,x_2)G(h)+H(h)\ ,\cr
\phi\sp 2(x_1,x_2)=&f_1(x_1,x_2)\ ,\cr
\phi\sp 3(x_1,x_2)=&f_2(x_1,x_2)\ ,\cr}\eqno(7.13)$$
where $f_1(x_1,x_2)$ and $f_2(x_1,x_2)$ as well as $F(h)$, $G(h)$ and $H(h)$
are arbitrary functions, and $h(x_1,x_2)$ is implicitly given by the relation
$$f_1(x_1,x_2){{\rm d}F\over{\rm d}h}(h)
+f_2(x_1,x_2){{\rm d}G\over{\rm d}h}(h)
+{{\rm d}H\over{\rm d}h}(h)=\ 0\ .\eqno(7.14)$$
Here again, the non degeneracy condition (4.4) requires that the Jacobian
$({\det({\pd f_j}/{\pd x_i})})$ is non vanishing, but it is easy to check
that the solutions to (1.3) which do not obey this condition are included
in (7.13) and (7.14).

Similarly, using the C-map, all solutions to the three dimensional Bateman
equation (1.2) are obtained from the implicit definitions
$$x_1F_1(h;\phi)+x_2F_2(h;\phi)+x_3F_3(h;\phi)+H(h;\phi)=\ 0\ ,\eqno(7.15)$$
and
$$x_1{\pd F_1\over\pd h}(h;\phi)+x_2{\pd F_2\over\pd h}(h;\phi)
+x_3{\pd F_3\over\pd h}(h;\phi)+{\pd H\over\pd h}(h;\phi)=\ 0\ .\eqno(7.16)$$
The functions $F_i(h;\phi)\ (i=1,2,3)$ and $H(h;\phi)$ are of course
arbitrary, and it is understood that (7.16) defines $h(x_1,x_2,x_3;\phi)$
which by
substitution in (7.15) then determines $\phi(x_1,x_2,x_3)$ implicitly. As in
the two dimensional case, the non degeneracy condition (3.6) excludes the
solutions for $\phi$ which are independent of $x_3$. However, it is easy to
show that the latter solutions as well as all those obtained from (7.11) and
(7.12) through the C-map may all be constructed from (7.15) and (7.16).
Note also how the equation (5.35), which defines implicit solutions to the
Bateman equation in any dimension, is clearly a particular reduction of
(7.15) and (7.16) in this case, corresponding to functions
$F_i(h;\phi)\ (i=1,2,3)$ and $H(h;\phi)$ which are independent of $h$.

Moreover, the expressions (7.15) and (7.16) also lead to an interesting
geometrical interpretation. Given a solution $\phi(x_1,x_2,x_3)$, consider
the equation $(\phi(x_1,x_2,x_3)=\phi_0)$ with $\phi_0$ being an arbitrary
constant. Such an equation defines a particular two dimensional
surface in the space parametrised by $(x_1,x_2,x_3)$. From (7.15) and (7.16),
one sees that this surface is the locus of the lines of
intersection of the two planes defined by these two conditions as the
parameter $h$ varies, or else, the surface is the plane defined by (7.15)
and (7.16) must
then be either trivial (corresponding to $F_i(h;\phi)\ (i=1,2,3)$ and
$H(h;\phi)$ independent of $h$) or proportional to (7.15) (corresponding to
$F_i(h;\phi)\ (i=1,2,3)$ and $H(h;\phi)$ being all proportional to the {\it
same} exponential function $({\rm exp}(\alpha(\phi)h))$\ ). Thus, the surface
defined by $(\phi(x_1,x_2,x_3)=\phi_0)$ is always a ruled surface
known$\sp {[11]}$ as a developable surface, namely a surface which may be
developed onto the plane. These surfaces are either planes, cones, cylinders
or tangent surfaces to curves ({\it i.e.} the locus of tangents to a curve).
Indeed, the characteristic function $(F(x_1,x_2,x_3)=0)$ for a developable
surface in three dimensions is precisely always a solution to the three
dimensional Bateman equation$\sp {[11]}$. Actually, by considering the
inverse C-map composed with the R-map, it may be seen that for any given
functions
$F_i(h;\phi)\ (i=1,2,3)$ and $H(h;\phi)$ in (7.15) and (7.16), the surfaces
defined by $(\phi(x_1,x_2,x_3)=\phi_0)$ and by (7.13) and (7.14) with
$(F(h)=-F_1(h;\phi_0)/F_3(h;\phi_0))$, $(G(h)=-F_2(h;\phi_0)/F_3(h;\phi_0))$
and $(H(h)=-H(h;\phi_0)/F_3(h;\phi_0))$ are the same surfaces. Thus, all
solutions to the universal string equations (1.3) in three spacetime
dimensions are$\sp {[2]}$ all developable surfaces. Note also that similar
considerations apply in the case $(d=1)$, developable surfaces being
then simply replaced by straight lines in a two dimensional space.

The results above have thus shown how it is possible in principle to find all
solutions to (1.1) -- (1.3) for $(d=0,1,2)$, thereby establishing their
implicit integrability. The situation for higher values of $d\ts$ is left as
an open question interesting in its own right. In this context, one may
wonder what type of hypersurface would generalise the r$\hat{\rm o}$le
played by developable surfaces above. Hereafter, general classes of
solutions to (1.1) -- (1.3) for arbitrary $d\ts$ are presented, but we do not
yet know how to construct, even implicitly, all solutions. However, the other
instance where the numbers of universal equations of motion and fields are
the same -- accounting for Ward identities in the reparametrisation invariant
case --, namely (6.6), (6.16) and (6.29), is very easily solved. The general
solution to (6.6) is of course
$$\phi\sp a(x)=A\sp a x+B\sp a\ ,\eqno(7.17)$$
with $A\sp a$ and $B\sp a$ being arbitrary constants. Using the C-map and
accounting for the degenerate cases not obeying (3.6), the general
solution to (6.16) is implicitly given by
$$x_iF_i\sp a(\phi\sp b)+H\sp a(\phi\sp b)=\ 0\ ,\eqno(7.18)$$
where $F_i\sp a(\phi\sp b)$ and $H\sp a(\phi\sp b)$ are arbitrary functions
of the fields.
Thus in this case, the constraints (1.6) define in fact {\it all} solutions
to the generalised Bateman equation (1.5) with $(D=d-1)$. Finally, using the
R-map and accounting for the condition (4.4), all solutions to (6.29) are
obtained as
$$\phi\sp a(x)=A\sp a f(x)+B\sp a\ ,\eqno(7.19)$$
where $f(x)$ is of course an arbitrary function.

Let us now present some classes of solutions to (1.1) -- (1.3) for arbitrary
$d\ts$. Obviously, whenever the field $\phi$ or one of the fields $\phi\sp a$
in case 3) are independent of at least one of the coordinates, we always
have a solution for either of the three equations. Consider now the generic
universal equation (1.1). A large class of solutions is given by
$$\phi(x_i)=G(x_i A_{in}+B_n)+H\ ,\eqno(7.20)$$
where $A_{in}$, $B_n$ and $H$ are arbitrary constants and $G(t_n)$ is an
arbitrary function of $N$ arguments $t_n$ with $(N < d\ts)$. In fact,
the function (7.20) also defines a solution when $(N=d\ts)$ provided either
the then square matrix $A_{in}$ or the matrix
$({\pd\sp 2 G}(t_n)/{\pd t_n\pd t_m})$ have a vanishing determinant (for all
values of $t_n$ in the latter case; any homogeneous function $G(t_n)$ of
weight one is such an example). Whatever the value of $N$, (7.20) is such
that $\phi(x_i)$ is in fact constant along at least one of the directions in
the space parametrised by the coordinates $x_i$. This situation is thus very
similar to the one described above concerning developable surfaces.

Given a specific solution to (1.1) in $d\ts$ dimensions, it is also possible
to construct iteratively solutions to (1.1) in higher dimensions. Namely, if
$\phi\sp {(d)}(x_i)$ is a solution in $d\ts$ dimensions, the function
defined by
$$\phi\sp {(d+1)}(x_i,x_{d+1})={\rm exp}\bigl[\phi\sp {(d)}(x_i)+\alpha x_{d+1}
+\beta\bigr]\ ,\eqno(7.21)$$
where $\alpha$ and $\beta$ are arbitrary constants, is always a solution to
(1.1) in $(d+1)$ dimensions. In particular, since all solutions to (1.1)
for $(d=2)$ are known, this procedure enables one to find large classes a
solutions in higher dimensions. Finally, we should also mention that using
the inverse C-map, it is possible to obtain still further solutions to (1.1)
from the solutions to the Bateman equation given hereafter.

Let us now turn$\sp {[1]}$ to the Bateman equation (1.2) in $d\ts$ dimensions.
As already mentioned, any field configuration obeying the relation (5.34),
thus in particular those functions defined implicitly by (5.35), always
provide solutions. Also, any function $\phi(x_i)$ homogeneous in the
variables $x_i$ and of zero weight, is always a solution. Finally, the
functions
$$\phi(x_i)=\prod_{n=1}\sp N\bigl[x_i A_{in}+B_n\bigr]\sp {\alpha_n}\ ,
\eqno(7.22)$$
where $A_{in}$, $B_n$ and $\alpha_n$ are arbitrary parameters, are always
solutions provided $(N < d\ts)$. If $(N=d\ts)$, we must have either
$({\det A_{in}}=0)$ or $(\sum_{n=1}\sp N\alpha_n=0)$.

Given a solution to (1.2) for a specific dimension, it is also
possible$\sp {[1]}$ to construct by iteration solutions in higher dimensions.
Namely, if $\phi\sp {(d)}(x_i)$ is a solution in $d\ts$ dimensions,
solutions in $(d+1)$ dimensions are obtained from
$$\phi\sp {(d+1)}(x_i,x_{d+1})=\phi\sp {(d)}(x_i+h_i(x_{d+1}))\ ,\eqno(7.23)$$
where $h_i(x_{d+1})$ are arbitrary functions. However, this recursive
procedure is different from the one obtained from (7.21) through the C-map
(the description of the latter is left to the reader). Thus, using the
C-map and its inverse, (7.21) and (7.23) provide two different iterative
procedures for constructing new solutions in higher dimensions from
specific ones in any particular dimension.

Finally, consider the C-map applied to the solutions in (7.20) for the
universal equation of the generic hierarchy. This leads to the following
implicit definition of solutions to the Bateman equation which are not
independent of $x_d$
$$x_d=G(x_\alpha A_{\alpha n}(\phi)+B_n(\phi))+H(\phi)\ .\eqno(7.24)$$
Here, $A_{\alpha n}(\phi)$, $B_n(\phi)$ and $H(\phi)$ are arbitrary
functions of $\phi$, $(\alpha=1,2,\cdots,d-1)$ and $G(t_n)$ are arbitrary
functions of $N$ arguments $t_n\ (n=1,2,\cdots,N)$ with
$(N < (d-1)\ts)$. When $(N=d-1\ts)$, we must have either
$({\det A_{\alpha n}}=0)$ or $({\det {\pd\sp 2 G}(t_n)/{\pd t_n\pd t_m}}=0)$
as discussed after (7.20). Clearly, the definition (7.24) includes (5.35)
as a special case, the latter corresponding to having $(N=1)$ and $G$ a
linear function of its single argument. Incidentally, the latter case, which
corresponds to applying the inverse C-map on (5.35), leads to the following
obvious solutions to (1.1)
$$\phi(x_i)=x_i A_i + B\ ,\eqno(7.25)$$
with $A_i$ and $B$ arbitrary constants. Consequently, we have
$$\phi_{ij}=\ 0\ ,\eqno(7.26)$$
which not only solves (1.1) most trivially, but also explains why (5.35)
always defines field configurations $\phi(x_i)$ which in fact trivially solve
{\it all} the equations of motion of the Bateman hierarchy, and not only
the Bateman equation itself -- a result pointed out after (5.35).

Let us conclude with some solutions to the generalised universal Bateman
equations (1.5), and their dual representation (6.5) in the generalised
generic hierarchy. Extending (5.34), field configurations obeying
identities of the form
$$\phi\sp a_{ij}=S\sp {ab}_i\phi\sp b_j+S\sp {ab}_j\phi\sp b_i\eqno(7.27)$$
for some coefficients $S\sp {ab}_i$, are always$\sp {[1]}$ solutions to (1.5).
In particular, this includes fields defined implicitly by the constraints
$$x_iF\sp a_i(\phi\sp b)+H\sp a(\phi\sp a)=0\ ,\eqno(7.28)$$
with $F\sp a_i(\phi\sp b)$ and $H\sp a(\phi\sp b)$ being arbitrary functions.
Such solutions are the generalisation of (5.35) already mentioned in the
introduction. Actually, it is easy to verify that configurations obeying
(7.27) also lead to
$$Y\sp a_{\alpha\beta}=\ 0\ ,\eqno(7.29)$$
with $Y\sp a_{\alpha\beta}$ defined in (3.9). This result shows that in fact
(7.27) defines solutions to {\it all} equations of motion of the generalised
Bateman hierarchy (6.10), and not only to the generalised Bateman equation
(6.15), thus extending the similar result for the Bateman hierarchy for one
field mentioned after (5.35).

It is also instructive to apply the inverse C-map to (7.28). One then obtains
the following trivial solutions to (6.5)
$$\phi\sp a(x_i)=x_i F\sp a_i+H\sp a\ ,\eqno(7.30)$$
with $F\sp a_i$ and $H\sp a$ now being arbitrary constants. Since we then have
$(\phi\sp a_{ij}=0)$, it is obvious why (7.28) always defines solutions to all
equations of motion for both the generalised generic and Bateman hierarchies
(6.2) and (6.10). Clearly, the same applies to (7.27) since
$Y\sp a_{\alpha\beta}$
then vanishes identically. In fact, in view of (7.20), (7.30) suggests an
obvious generalisation. Namely, consider the fields
$$\phi\sp a(x_i)=G\sp a(x_iA_{in}+B_n)+H\sp a\ ,\eqno(7.31)$$
where $A_{in}$, $B_n$ and $H\sp a$ are arbitrary coefficients and
$G\sp a(t_n)$ are
arbitrary functions of $N$ arguments $t_n$ with $(N < d\ts)$. If $(N=d\ts)$,
we must also have $({\det A_{in}}=0)$. These field configurations, clearly
generalising (7.30), are always solutions to the universal
equations of motion (6.5). Using the C-map, one then also obtains the
following constraints defining implicitly solutions to the generalised Bateman
equations (6.15)
$$y\sp a=G\sp a(x_\alpha A_{\alpha n}(\phi\sp b)+B_n(\phi\sp b))
+H\sp a(\phi\sp b)\ .\eqno(7.32)$$
Here, $A_{\alpha n}(\phi\sp b)$, $B_n(\phi\sp b)$, $H\sp a(\phi\sp b)$
and $G\sp a(t_n)$ are
arbitrary functions with $(n=1,2,\cdots,N)$ and $(N < (d-D)\ts)$.
If $(N=d-D\ts)$, we must also have $({\det A_{\alpha n}}(\phi\sp b)=0)$ (the
coordinates $x_i$ have been split into $q\ts$ coordinates $x_\alpha$ and
$p\ts$ coordinates $y\sp a$, where $p\ts$ is the number of fields
$\phi\sp a$ and
$(d=p+q)$; see sect.3)). The solutions to the generalised Bateman equation
(1.5) defined by the constraints (7.32) extend those provided by (7.28) but
do not obey (7.27), so that such configurations are no longer solutions to
all the other equations of motion of the generalised Bateman hierarchy.
Finally, using the R-map, it should be clear how (7.31) also defines
solutions to the reparametrisation invariant universal string and membrane
equations of motion (6.27). As in (7.20), one then finds string and
membrane configurations which are constant along at least one of the
directions in spacetime, a situation which is not avoided even when
implementing the iterative procedure (7.21). This class of solutions is a
straightforward extension of developable surfaces to higher dimensions.
However, it is not clear at all whether {\it all} solutions to the
generalised universal equations of this paper would be of this type.
\vskip 20pt
\leftline{\bf 8. Conclusions}
\vskip 20pt
The results of this paper have used two main ingredients. On the one hand,
a triangular duality relationship between certain classes of arbitrary field
theories, of classical topological field theories and of new types of string
and membrane theories. On the other hand, the existence of a generic finite
Euler hierarchy of Lagrangians and their equations of motion leading to a
universal equation for one field in arbitrary dimensions which generalises
the Plebanski equation$\sp {[3]}$ for self-dual gravity in four dimensions.

The duality maps have made explicit the close similarity between
reparametrisation invariant theories such as strings and membranes, and
classical topological field theories whose space of classical solutions falls
into diffeomorphic topological classes of the target manifold. The generic
hierarchy has shown that there is a close connection -- not yet properly
understood -- between the finiteness of a
Euler hierarchy, the universality of its last non trivial equations of
motion, and the fact that these equations of motion, even though non linear in
the fields, are multilinear in their second derivatives and of order equal to
the level at which they arise in the hierarchy.

The universal equations of motion have the peculiarity that they derive from
an infinity of inequivalent Lagrangians and are covariant under
arbitrary linear transformations both in base space and in target space, even
though the associated Lagrangians do not necessarily possess these symmetries.
Moreover, the infinity of Lagrangians sharing the same equations of motion
strongly suggests the possible integrability of the universal equations of
motion. This is indeed confirmed in the simplest cases where the complete
solution to the equations can be specified using simple methods.

These results obviously raise a series of interesting questions. Clearly, one
may wonder whether there exist finite Euler hierarchies leading to
universal equations of motion other than the generic hierarchy and all its
descendants constructed here. Could such hierarchies also exist
for anticommuting fields? Is the supersymmetrisation of the present
hierarchies possible while preserving their finiteness and universality? Do
such finite Euler hierarchies exhaust all classes of field equations admitting
an infinity of inequivalent Lagrangians? In any
case, were such new hierarchies to exist, one would also
obtain their dual representations as new classical topological field
theories and new string and membrane theories.

The question of integrability of the universal equations of this paper -- and
of any others were they to exist -- also begs for an answer, which may well
turn out to have profound consequences for our understanding of integrable
systems in higher dimensions.

Of course, there is also the whole problem of quantising the infinity of field
theories leading to these universal equations of motion. In the context of
their ``Bateman representation'' as classical topological field theories, it
would certainly be of interest to find out what becomes of the classical
solutions once these systems are quantised, since such systems possess a
novel kind of gauge symmetry only realised on the space of classical
solutions but not at the level of the action. In the context of the string
or membrane
representation, the whole issue of a consistent BRST quantisation$\sp {[10]}$
of these gauge invariant theories certainly warrants a close examination,
having in mind among other issues the possibility of critical dimensions
and the question of physical states.

On the other hand, the generic hierarchy uses the Euler operator ${\cal E}$
as an essential ingredient in its construction. In fact, this operator is
nilpotent, {\it i.e.} $({\cal E}\sp 2=0)$, when acting on Lagrangians having
a  dependence only on derivatives of fields but not on the fields themselves,
as is the case here. In other words, this operator is in our case$\sp {[1]}$
in the nature of the exterior derivative in differential geometry or the
BRST charge in gauge invariant systems. It is not clear whether the
associated cohomological structure inherent$\sp {[1]}$ to the construction
of Euler hierarchies plays any crucial r$\hat{\rm o}$le
in the existence of the generic finite
hierarchy. Nevertheless, it certainly raises the question whether other finite
Euler hierarchies could not be constructed on the basis of a modified Euler
operator more in the nature of a {\it covariant} exterior derivative for some
principal fiber bundle or some generalisation thereof in the context of
infinite dimensional manifolds of field configurations.

Finally, there is also the issue of the possible physical relevance of these
universal equations. The obvious suggestion is in the context of theories for
gravity. On the one hand, the fields of a classical topological field theory
may be thought of as spacetime coordinates, since classical solutions
are covariant under arbitrary field redefinitions. In this respect, the
quantisation of these theories becomes more pressing. On the other hand,
in view of the situation with ordinary critical strings based on the
Nambu-Goto action and its different supersymmetric extensions,
reparametrisation invariant universal equations could
possibly also be of relevance to theories of gravity. Here again, the
understanding of the space of physical states at the quantum level is
required. However, one may also wonder whether the universal equations
of this paper have any other connection with present
developments of string theory, either in the context of the geometrical
realisation of W-algebras$\sp {[12]}$ or indeed of W-gravity
theories$\sp {[13]}$,
or in the context of the matrix model formulation$\sp {[14,15,16]}$ of two
dimensional quantum gravity and non critical string theories, this latter
formulation being directly related to two dimensional integrable systems
such as the KdV hierarchy which indeed made its appearance here as well.

\vskip 20pt
\leftline{\bf Acknowledgement}
\vskip 10pt
The work of J.G. is supported through a Senior Research Assistant position
funded by the S.E.R.C.

\vfill\eject
\vskip 10pt
\leftline{\bf Appendix}
\vskip 20pt
In complement to Sect.5, giving the proof for the generic finite Euler
hierarchy for one field, definitions and some properties of generalised
determinant and trace functions of matrices are collected in this Appendix.
These generalised determinants are multilinear functions of collections of
arbitrary $d\times d$ matrices
$A\sp {(p)}_{ij}\ (p=1,2,\ldots\ ;\ i,j=1,2,\cdots,d\ts)$. These functions
include as particular cases the ordinary determinant and trace functions
when all matrices $A\sp {(p)}_{ij}$ are identical.

First, for any $(0\leq n\leq d\ts)$, let us introduce the following
quantities
$$\eqalign{T\sp {(n)}_{i_1\cdots i_n;j_1\cdots j_n}=&
\sum_{\sigma_n}\vert\sigma_n\vert\delta_{i_1\sigma_n(j_1)}\cdots
\delta_{i_n\sigma_n(j_n)}\cr
=&\ {1\over (d-n)!}\ \epsilon_{i_1\cdots i_d}\epsilon_{j_1\cdots j_d}
\delta_{i_{n+1}j_{n+1}}\cdots\delta_{i_d j_d}\ .}\eqno(A.1)$$
In this definition, the summation is over all permutations $\sigma_n$ of
$n\ts$ elements, and $\vert\sigma_n\vert$ denotes the signature of each
permutation. In particular, we have $T\sp {(0)}=1$. Note that the quantities
$T\sp {(n)}_{i_1\cdots i_n;j_1\cdots j_n}$ are symmetric in the n-plets
$(i_1\cdots i_n)$ and $(j_1\cdots j_n)$ and antisymmetric in the
indices $i_k$ and $j_k$ separately. They also obey the identities
$$\delta_{i_{n+1}j_{n+1}}T\sp {(n+1)}_{i_1\cdots i_{n+1};j_1\cdots j_{n+1}}=
(d-n)\ T\sp {(n)}_{i_1\cdots i_n;j_1\cdots j_n}\ ,\eqno(A.2)$$
and
$$T\sp {(n+1)}_{i_1\cdots i_{n+1};j_1\cdots j_{n+1}}=
T\sp {(n)}_{i_1\cdots i_n;j_1\cdots j_n}\delta_{i_{n+1}j_{n+1}}-
\sum_{p=1}\sp nT\sp {(n)}_{i_1\cdots i_n;
j_1\cdots j_{p-1}j_{n+1}j_{p+1}\cdots j_n}
\delta_{i_{n+1}j_p}\ .\eqno(A.3)$$
The second of these relations, which is the converse of the first, is the
combinatorial property at the origin of the generic hierarchy of Sect.5.

Given a collection of arbitrary matrices $A\sp {(p)}_{ij}\ (p=1,2,\cdots,n)$
with
$(0\leq n\leq d\ts)$, consider now the multilinear functions defined by
$$L\sp {(n)}_{i_{n+1}\cdots i_d;j_{n+1}\cdots j_d}
(A\sp {(1)},\cdots,A\sp {(n)})=
{1\over n!(d-n)!}\ \epsilon_{i_1\cdots i_d}\ \epsilon_{j_1\cdots j_d}
A\sp {(1)}_{i_1 j_1}\cdots A\sp {(n)}_{i_n j_n}\ ,\eqno(A.4)$$
and
$$\eqalign{Tr\sp {(n)}(A\sp {(1)},\cdots,A\sp {(n)})=&
\ L\sp {(n)}_{i_{n+1}\cdots i_d;i_{n+1}\cdots i_d}
(A\sp {(1)},\cdots,A\sp {(n)})\cr
=&\ {1\over n!}\ T\sp {(n)}_{i_1\cdots i_n;j_1\cdots j_n}
A\sp {(1)}_{i_1 j_1}\cdots A\sp {(n)}_{i_n j_n}\ .\cr}\eqno(A.5)$$
When all arguments $A\sp {(p)}$ are identical, these functions will be
denoted $L\sp {(n)}_{i_{n+1}\cdots i_d;j_{n+1}\cdots j_d}(A)$ and
$Tr\sp {(n)}(A)$ respectively. Note that we have
$$Tr\sp {(1)}(A)={\rm tr}\ A\ ,\quad
Tr\sp {(d)}(A)=\det A=L\sp {(d)}(A)\ ,\eqno(A.6)$$
showing that the functions (A.4) and (A.5) indeed generalise the ordinary
trace and determinant functions of matrices. In fact, the functions
$L\sp {(n)}_{i_{n+1}\cdots i_d;j_{n+1}\cdots j_d}(A)$
essentially correspond to the minors of order $n\ts$ of the matrix $A$.

Clearly, the functions $L\sp {(n)}$ and $Tr\sp {(n)}$ are fully symmetric
functions of their arguments $A\sp {(p)}\ (p=1,2,\cdots,n\ts)$. Also, the
functions
$L\sp {(n)}_{i_{n+1}\cdots i_d;j_{n+1}\cdots j_d}
(A\sp {(1)},\cdots,A\sp {(n)})$ are
antisymmetric under the exchange of any pair of the $i_k$ or of the $j_k$
indices. Moreover, these functions are symmetric under the exchange of the
$(d-n)$-plets $(i_{n+1}\cdots i_d)$ and $(j_{n+1}\cdots j_d)$ only if
{\it all} the arguments $A\sp {(p)}\ (p=1,2,\cdots,n\ts)$ are symmetric
matrices. Finally, we have the following relations and particular cases
$$\eqalignno{L\sp {(n)}_{i_{n+1}\cdots i_d;j_{n+1}\cdots j_d}
({A\sp {(1)}}\sp {\rm T},\cdots,{A\sp {(n)}}\sp {\rm T})=&
\ L\sp {(n)}_{j_{n+1}\cdots j_d;i_{n+1}\cdots i_d}
(A\sp {(1)},\cdots,A\sp {(n)})\ ,&(A.7)\cr
Tr\sp {(n)}({A\sp {(1)}}\sp {\rm T},\cdots,{A\sp {(n)}}\sp {\rm T})=&
\ Tr\sp {(n)}(A\sp {(1)},\cdots,A\sp {(n)})\ ,&(A.8)\cr
L\sp {(0)}_{i_1\cdots i_d;j_1\cdots j_d}=
{1\over d!}\ \epsilon_{i_1\cdots i_d}&\ \epsilon_{j_1\cdots j_d}
={1\over d!}\ T\sp {(d)}_{i_1\cdots i_d;j_1\cdots j_d}\ ,&(A.9)\cr
L\sp {(d)}(A\sp {(1)},\cdots,A\sp {(d)})=&\ Tr\sp {(d)}
(A\sp {(1)},\cdots,A\sp {(d)})\ .&(A.10)\cr}$$

Generalising the expansion of the ordinary determinant function in terms of
minors of a matrix, we also have the following identities for any $p<n$
$$\eqalign{L\sp {(n)}_{i_{n+1}\cdots i_d;j_{n+1}\cdots j_d}&
(A\sp {(1)},\cdots,A\sp {(n)})=\cr
&={p!(d-p)!\over n!(d-n)!}
\ L\sp {(p)}_{i_{p+1}\cdots i_d;j_{p+1}\cdots j_d}
(A\sp {(1)},\cdots,A\sp {(p)})
A\sp {(p+1)}_{i_{p+1}j_{p+1}}\cdots A\sp {(n)}_{i_n j_n}\ ,\cr}\eqno(A.11)$$
and
$$\eqalign{Tr\sp {(n)}&(A\sp {(1)},\cdots,A\sp {(n)})=\cr
&={p!(d-p)!\over n!(d-n)!}
\ L\sp {(p)}_{i_{p+1}\cdots i_d;j_{p+1}\cdots j_d}
(A\sp {(1)},\cdots,A\sp {(p)})
A\sp {(p+1)}_{i_{p+1}j_{p+1}}\cdots A\sp {(n)}_{i_n j_n}
\delta_{i_{n+1}j_{n+1}}\cdots\delta_{i_d j_d}\ .}\eqno(A.12)$$
In particular, since $Tr\sp {(d)}(A)=\det A=L\sp {(d)}(A)$,
the last identity shows that
$$L\sp {(d-1)}_{i;j}(A)=\widetilde A_{ji}\ ,\eqno(A.13)$$
where $\widetilde A_{ij}$ is the adjugate matrix of $A_{ij}$ given by
$$\widetilde A=(\det A)\ A\sp {-1}\ .\eqno(A.14)$$

Finally, generalising the well known result for ordinary determinants,
we have the multiplication theorems
$$\eqalignno{L\sp {(n)}_{i_{n+1}\cdots i_d}&_{;j_{n+1}\cdots j_d}
(A\sp {(1)}B,\cdots,A\sp {(n)}B)=\cr
&=L\sp {(n)}_{i_{n+1}\cdots i_d;k_{n+1}\cdots k_d}
(A\sp {(1)},\cdots,A\sp {(n)})
\ L\sp {(n)}_{k_{n+1}\cdots k_d;j_{n+1}\cdots j_d}(B)\ ,&(A.15a)\cr
L\sp {(n)}_{i_{n+1}\cdots i_d}&_{;j_{n+1}\cdots j_d}
(BA\sp {(1)},\cdots,BA\sp {(n)})=\cr
&=L\sp {(n)}_{i_{n+1}\cdots i_d;k_{n+1}\cdots k_d}(B)
\ L\sp {(n)}_{k_{n+1}\cdots k_d;j_{n+1}\cdots j_d}
(A\sp {(1)},\cdots,A\sp {(n)})\ ,&(A.15b)}$$
and
$$\eqalign{Tr\sp {(n)}&(A\sp {(1)}B,\cdots,A\sp {(n)}B)=
Tr\sp {(n)}(BA\sp {(1)},\cdots,BA\sp {(n)})=\cr
&=L\sp {(n)}_{i_{n+1}\cdots i_d;j_{n+1}\cdots j_d}
(A\sp {(1)},\cdots,A\sp {(n)})
\ L\sp {(n)}_{j_{n+1}\cdots j_d;i_{n+1}\cdots i_d}(B)\ .}\eqno(A.16)$$
In particular, we thus obtain
$$\eqalign{Tr\sp {(d)}(A\sp {(1)}B,\cdots,A\sp {(d)}B)&=
Tr\sp {(d)}(BA\sp {(1)},\cdots,BA\sp {(d)})=\cr
&=(\det B)\ Tr\sp {(d)}(A\sp {(1)},\cdots,A\sp {(d)})\ ,\cr}\eqno(A.17)$$
and
$$\eqalign{Tr\sp {(d-1)}(A\sp {(1)}B,\cdots,A\sp {(d-1)}B)&=
Tr\sp {(d-1)}(BA\sp {(1)},\cdots,BA\sp {(d-1)})=\cr
&=\widetilde B_{ij}\ L\sp {(d-1)}_{i;j}
(A\sp {(1)},\cdots,A\sp {(d-1)})\ ,\cr}\eqno(A.18)$$
where $\widetilde B$ is the adjugate of $B$ (see (A.14)). Note that these
results, which are rather easy to prove, require that the {\it same}
matrice $B$ multiplies the matrices $A\sp {(p)}$.

As we have seen, the functions $Tr\sp {(d)}=L\sp {(d)}$ generalise the
ordinary
determinant of a $d\times d$ matrix to the case of $d\ts$ such matrices.
As is well known, the determinant of a matrix vanishes whenever at least one
of its lines (or columns) is a linear combination of the other lines (or
columns). More generally here, we have that
$$Tr\sp {(d)}(A\sp {(1)},\cdots,A\sp {(d)})=
L\sp {(d)}(A\sp {(1)},\cdots,A\sp {(d)})=0\ ,\eqno(A.19)$$
whenever {\it all} matrices $A\sp {(p)}\ (p=1,2,\cdots,d\ts)$ have a
vanishing determinant and for each one of them, the {\it same} linear
combination of its lines (or columns), {\it with the same coefficients for
all matrices}, applies. The converse is also true, at least when all
matrices which are different among the set of $d\ts$ matrices are also
linearly independent.

The proof of this statement is straightforward. For definiteness, let us
assume that it is the last column of each matrix which is a linear
combination of its other columns, namely
$$A\sp {(p)}_{id}=\sum_{\beta =1}\sp {d-1}A\sp {(p)}_{i\beta}
C_\beta\ ,\qquad{\rm for}\quad p=1,2,\cdots,d\ .\eqno(A.20)$$
The coefficients $C_\beta$ are thus the {\it same} for all matrices
$A\sp {(p)}$ ($\alpha ,\beta=1,2,\cdots,(d-1)$). Given (A.20),
we actually have
$$A\sp {(p)}_{ij}=\pmatrix{A\sp {(p)}_{\alpha\beta}&
A\sp {(p)}_{\alpha\beta}C_\beta\cr
A\sp {(p)}_{d\beta}&A\sp {(p)}_{d\beta}C_\beta\cr}=
\pmatrix{A\sp {(p)}_{\alpha\beta}&0\cr A\sp {(p)}_{d\beta}&0\cr}
\pmatrix{\delta_{\alpha\beta}&C_\alpha\cr0&0\cr}\ .\eqno(A.21)$$
Here, $i$ or $\alpha$ and $j$ or $\beta$ denote line and column indices
respectively. Using the multiplication theorem (A.17), it is clear that
(A.19) is indeed obtained.

To conclude, let us now consider matrices of the form
$$A\sp {(p)}_{ij}=
\pmatrix{A\sp {(p)}_{\alpha\beta}&A\sp {(p)}_{\alpha\beta}C_\beta\cr
C_\alpha A\sp {(p)}_{\alpha\beta}&C_\alpha
A\sp {(p)}_{\alpha\beta}C_\beta\cr}\ ,\eqno(A.22)$$
where $C_\alpha$ are arbitrary coefficients common to {\it all}
matrices $A\sp {(p)}_{ij}\ (p=1,2,\cdots,d\ts)$ (the notation is the same
as in (A.20)). From the previous result, we know that the functions
$Tr\sp {(d)}=L\sp {(d)}$ then vanish for these matrices. However, the
functions $Tr\sp {(d-1)}$ and $L\sp {(d-1)}_{i;j}$ need not vanish
-- in the same way that the minors of order $(d-1)$ of a $d\times d$
matrix of vanishing determinant need not vanish. Given (A.22),
we have in fact
$$A\sp {(p)}_{ij}=\pmatrix{\delta_{\alpha\beta}&0\cr C_\beta&0\cr}
\pmatrix{A\sp {(p)}_{\alpha\beta}&0\cr 0&0\cr}
\pmatrix{\delta_{\alpha\beta}&C_\alpha\cr 0&0\cr}\ .\eqno(A.23)$$
{}From a straightforward calculation, one now obtains
$$\eqalign{&L\sp {(d-1)}_{i;j}
(\pmatrix{\delta_{\alpha\beta}&0\cr C_\beta&0\cr})=
\pmatrix{0&-C_\alpha\cr 0&1\cr}\ ,\cr
&L\sp {(d-1)}_{i;j}(\pmatrix{\delta_{\alpha\beta}&C_\alpha\cr 0&0\cr})=
\pmatrix{0&0\cr -C_\beta&1\cr}\ ,\cr
&L\sp {(d-1)}_{i;j}(\pmatrix{A\sp {(1)}_{\alpha\beta}&0\cr 0&0\cr},\cdots,
\pmatrix{A\sp {(d-1)}_{\alpha\beta}&0\cr 0&0\cr})=
Tr\sp {(d-1)}_{(d-1)}(A\sp {(1)}_{\alpha\beta},\cdots,
A\sp {(d-1)}_{\alpha\beta})
\pmatrix{0&0\cr 0&1\cr}\ .}\eqno(A.24)$$
In the last equality, the function $Tr\sp {(d-1)}_{(d-1)}$ is a generalised
determinant for $(d-1)\times (d-1)$ matrices. From these results, (A.23) and
the multiplication theorem (A.15), one finally concludes that for matrices of
the form (A.22), the functions $L\sp {(d-1)}_{i;j}$ take the following values
$$L\sp {(d-1)}_{i;j}(A\sp {(1)},\cdots,A\sp {(d-1)})=
Tr\sp {(d-1)}_{(d-1)}(A\sp {(1)}_{\alpha\beta},\cdots,
A\sp {(d-1)}_{\alpha\beta})
\pmatrix{C_\alpha C_\beta& -C_\alpha\cr -C_\beta&1\cr}\ .\eqno(A.25)$$

\vfill\eject
\leftline{\bf REFERENCES}
\vskip 20pt
\frenchspacing
\item{[1]}D. B. Fairlie, J. Govaerts and A. Morozov, {\sl Universal Field
Equations with Covariant Solutions}, Durham preprint DTP-91/55,
hepth-9110022 (October 1991), published in {\it Nuclear Physics B\/}.
\item{[2]}D. B. Fairlie and J. Govaerts, {\sl Universal Field Equations with
Reparametrisation Invariance}, Durham preprint DTP-92/11, hepth-9202056
(February 1992), to appear in {\it Physics Letters B\/}.
\item{[3]}J. D. Finley III and J. F. Plebanski, {\it Jour. Math. Phys.}
{\bf 17} (1976) 585.
\item{[4]}H. Bateman, {\it Proc. Roy. Soc. London} {\bf A 125} (1929) 598.
\item{[5]}P. R. Garabedian, {\it Partial Differential Equations}
(John Wiley \& Sons, New York, 1964) p. 517.
\item{[6]}E. Witten, {\it Comm. Math. Phys.} {\bf 117} (1988) 353;
{\it ibid.} {\bf 118} (1988) 411.
\item{[7]}Y. Nambu, {\it Lectures at the Copenhagen Symposium} (1970).
\item{[8]}T. Goto, {\it Prog. Theor. Physics} {\bf 46} (1971) 1560.
\item{[9]}E. Olver, {\it Applications of Lie Groups to Differential
Equations}, Graduate Texts in Mathematics {\bf 107} (Springer Verlag,
Berlin, 1986) p. 252.
\item{[10]}For a recent review, see \hfil\break
J. Govaerts, {\it Hamiltonian Quantisation and Constrained Dynamics},
Lecture Notes in Mathematical and Theoretical Physics {\bf 4}
(Leuven University Press, Leuven, 1991).
\item{[11]}L. P. Eisenhart, {\it An Introduction to Differential Geometry},
(Princeton University Press, Princeton, 1940) p. 59.
\item{[12]}J.-L. Gervais and Y. Matsuo, {\it Phys. Lett.} {\bf B274}
(1992) 309; preprint LPTENS-91/35, NBI-HE-91/50, hepth-9201026 (January 1992).
\item{[13]}C. M. Hull, {\it Phys. Lett.} {\bf B269} (1991) 257.
\item{[14]}E. Brezin and V. A. Kazakov, {\it Phys. Lett.} {\bf B236}
(1990) 144.
\item{[15]}M. Douglas and S. Shenker, {\it Nucl. Phys.} {\bf B335}
(1990) 635.
\item{[16]}D. Gross and A. A. Migdal, {\it Phys. Rev. Lett.} {\bf 64}
(1990) 127; {\it Nucl. Phys.} {\bf B340} (1990) 333.
\end